\documentclass[%
 reprint,
superscriptaddress,
 amsmath,amssymb,
 aps,
pra,
]{revtex4-1}
\usepackage{color}
\usepackage{xcolor}
\usepackage{physics}
\usepackage[normalem]{ulem}

\usepackage{graphicx}% Include figure files
\newcommand{\nn}{\nonumber}
\usepackage{dcolumn}% Align table columns on decimal point
\usepackage{bm}% bold math
\usepackage{amsthm}
\newcommand{\cu}[1]{\left\{ {#1} \right\}}
\newcommand{\ro}[1]{\left( {#1} \right)}
\usepackage{hyperref}
\hypersetup{hypertex=true,
colorlinks          =true,
linkcolor           =blue,
anchorcolor         =blue,
citecolor           =blue} 
\usepackage[mathlines]{lineno}% Enable 

\begin{document}

\preprint{APS/123-QED}

\title{Robust one-sided self-testing of two-qubit states via quantum steering}% 

\author{Yukun Wang }
 \affiliation{Beijing Key Laboratory of Petroleum Data Mining, China University of Petroleum, Beijing 102249, China}
\affiliation{State Key Laboratory of Cryptology, P.O. Box 5159, Beijing, 100878, China}

\author{Xinjian Liu}
 \affiliation{Beijing Key Laboratory of Petroleum Data Mining, China University of Petroleum, Beijing 102249, China}
 
\author{Shaoxuan Wang}
 \affiliation{Beijing Key Laboratory of Petroleum Data Mining, China University of Petroleum, Beijing 102249, China}
 
\author{Haoying Zhang}
 \affiliation{Beijing Key Laboratory of Petroleum Data Mining, China University of Petroleum, Beijing 102249, China}

\author{Yunguang Han}%
\email{hanyunguang@nuaa.edu.cn}
\affiliation{College of Computer Science and Technology, Nanjing University of Aeronautics and Astronautics, Nanjing 211106, China}

\date{\today}% 

\begin{abstract}
Entangled two-qubit states are the core building blocks for constructing quantum communication networks.
Their accurate verification is crucial to the functioning of the networks, especially for untrusted networks.
In this work we study the self-testing of two-qubit entangled states via steering inequalities, with robustness
analysis against noise. More precisely, steering inequalities are constructed from the tilted Clauser-Horne-Shimony-Holt inequality and its general form, to verify the general two-qubit entangled states. The study
provides a good robustness bound, using both local extraction map and numerical semidefinite-programming
methods. In particular, optimal local extraction maps are constructed in the analytical method, which yields the
theoretical optimal robustness bound. To further improve the robustness of one-sided self-testing, we propose a
family of three measurement settings steering inequalities. The result shows that three-setting steering inequality
demonstrates an advantage over two-setting steering inequality on robust self-testing with noise. Moreover, to
construct a practical verification protocol, we clarify the sample efficiency of our protocols in the one-sided
device-independent scenario.

\begin{description}
\item[Usage]
Secondary publications and information retrieval purposes.

\end{description}
\end{abstract}

\maketitle

\section{Introduction}

Quantum entangled states is the key resource of quantum information technologies, such as quantum networks \cite{Kimble2008}, cryptography \cite{Xu2020}, computation \cite{Campbell2017}, and metrology \cite{Giovannetti2011}. As we advance towards the second quantum revolution \cite{Deutsch2020}, the  characterization and certification of quantum devices becomes an extremely important topic in the practical applications of quantum technologies \cite{Eisert2020,Kliesch2021}. 

To ensure the proper functioning of a quantum network, it is essential to certify the entangled state deployed in the network accurately and efficiently. Besides the traditional quantum state tomography method, various methods have been proposed to improve the efficiency and apply to different scenarios, such as direct fidelity estimation \cite{Flammia2011}, compressed sensing tomography \cite{Gross2010}, and shadow tomography \cite{Huang2020}. In the last few years, quantum state verification (QSV) has attracted much attention by achieving remarkably low sample efficiency \cite{Pallister2018,Zhu2019}. One drawback of quantum state verification method is that it requires the perfect characterization of  the measurements performed by the quantum devices, thus it is device dependent and not applicable to the untrusted quantum network. Self-testing \cite{Supic2020,Mayers2004} is a prominent candidate of quantum state certification in device-independent (DI) scenario, in which all quantum devices are treated as black-boxes. Taking the advantage of Bell nonlocality \cite{Nonlocality2014}, many important results on self-testing have been achieved, such as self-testing various quantum entangled states \cite{McKague2012,Yang2014,Coladangelo2017}, self-testing entangled quantum measurement \cite{Renou2018,Bancal2018}, and parallel self-testing \cite{Reichardt2013,Wu2016}. Self-testing has wide applications in device-independent quantum information tasks, such as device-independent quantum random number generation \cite{Pironio2010,Liu2018}, and quantum key distribution \cite{Acin2007,Vazirani2014}.

Lying between standard QSV and self-testing, there is semi-device-independent (SDI) scenario \cite{Shrotriya2021} in which some parties are honest, while some others may be dishonest. The certification in this scenario can be called as SDI self-testing or SDI state verification. This scenario has wide applications in quantum information processing, such as one-sided device-independent (1SDI) quantum key distribution \cite{Branciard2012}, quantum random number generation \cite{Passaro2015}, verifiable quantum computation \cite{Gheorghiu2017}, and anonymous communication \cite{Unnikrishnan2019,Hahn2020,Wang2022}. Meanwhile the certification in the SDI scenario is closely related to the foundational studies on quantum steering in the untrusted quantum networks \cite{Uola2020,Wiseman2007,Saunders2010,Cavalcanti2015a}. However, not much is known about the quantum certification in the SDI scenario despite its significance. In \cite{Gheorghiu2017,Supic2016}, the authors studied the one-sided self-testing of maximally entangled two-qubit state based on 2-setting quantum steering inequality. In \cite{Han2021}, the authors proposed various verification protocols for Bell state based on multiple settings. For nonmaximal entangled two-qubit states, the authors in \cite{Goswami2018} realized the one-sided certification by combining fine-grained inequality \cite{Pramanik2014} and analog CHSH inequalities \cite{Cavalcanti2015}, which is more complicated compared with traditional self-testing.  In \cite{Shrotriya2021}, the authors proposed tilted steering inequality analogous to  tilted-CHSH inequality \cite{Acin2012} for one-sided self-testing of two-qubit states.  Then they generalized the  one-sided certification to general pure bipartite states by adopting the subspace method in DI scenario \cite{Coladangelo2017}. {In Ref. \cite{Sarkar2021}, 
a class of steering inequalities concentrating on the nonmaximal  entangled bipartite-qudit state were constructed, where they achieve the bipartite-qudit state self-testing by performing only two measurements. While in Ref. \cite{Skrzypczyk2018}, steering inequalities with $d+1$ measurement settings are used for self-testing the same states.} However, the robustness analysis there follows the norm inequalities method in \cite{McKague2012,Supic2016} (if it's not missed), thus the result is quite weak.  For the multipartite case, the studies of SDI certification are mainly focused on Greenberger–Horne–Zeilinger (GHZ) states as the generalization of Bell state \cite{Pappa2012,McCutcheon2016,Han2021}.

In this paper, we focus on the robust one-sided self-testing of two-qubit entangled states. We construct two types of 2-setting steering inequalities for general two-qubit entangled states based on tilted-CHSH inequality and its general form. For the first type, analytical and optimal robustness bound is obtained using the local extraction channel method introduced in \cite{Kaniewski2016}. For the second type, we get nearly linear robustness bound using numerical method based on the swap trick \cite{Yang2014} and semidefinite programming (SDP). To put our work in perspective, we
compare the robustness result in the 1SDI scenario with both DI and device-dependent scenario. Our result can be applied to the certification of high dimensional quantum devices as building blocks.  

Furthermore, we construct three measurement settings steering inequalities for general two-qubit states, which is beyond the conventional one-sided self-testing based on two settings. In \cite{Han2021}, the authors studied the optimal verification of Bell state and GHZ states in the 1SDI scenario using multiple measurement settings. However, their study is limited to the maximal entangled state in bipartite case. Based on the 3-settings steering inequalities, it is shown that the robustness bound can be further improved. This opens the question that how much the resistance to noise can be improved using multiple measurement settings.  Finally, to construct a practical verification protocol, we clarify the sample efficiency for our protocols in the 1SDI scenario. It is shown that approximately optimal sample efficiency can be obtained based on the steering inequalities we constructed.

\section{Preliminary}

\subsection{\label{sec:steeringscenario}  Steering scenario and steering inequalities}

Let us start by recalling the steering theory. Two distant parties, Alice and Bob, are considered, and between them are many copies of 
state $\rho_{AB}\in H_A\bigotimes H_B$. Bob performs
two  measurements labeled by $y$, on his particle and
obtains the binary outcome $b$. Meanwhile, Alice receives the corresponding
unnormalized conditional states $\rho_{b|y}$ and performs measurements randomly, labeled by $x$, and
obtains the binary outcome $a$. If Alice cannot explain
the assemblage of received states by assuming pre-existing states at her location and some pre-shared random numbers with Bob, she has to believe that Bob has the steerability of her particle from a distance.
 To determine whether Bob has steerability of her,  Alice asks Bob to run the experiment many times with her.  Finally, they  obtain the measurement statistics.
If the statistics admit the description, 
\begin{align}
 p(a, b|x, y; \rho_{AB}) = \sum\limits_{\lambda} p(\lambda)p(a|x,\rho_{\lambda})p(b|y,\lambda), 
\end{align}
then Alice knows Bob has not the steerability of her. This  non-steerable correlation models is  the so called local hidden variable (LHV)-LHS model \cite{Cavalcanti2015}.
The LHV-LHS decomposition is based on the idea that
Bob’s outcomes are determined by a local hidden random $\lambda$
and Alice’s outcomes are determined by local measurements
on quantum state $\rho_\lambda$.

 The  combination of the statistics will give a
steering inequality, where the LHV-LHS model can be used to
establish local bounds for the
steering inequality;  violation of such inequalities implies
steering. In Ref. \cite{Saunders2010}, the authors introduced a family of steering inequalities for Bell state
\begin{align}\label{eq:Bellsteering}
S_n\equiv \frac{1}{n}\sum_{k=1}^n \langle\hat{\sigma}_k^A{B_k} \rangle\leq C_n,
\end{align}
 $C_n$ is the LHS bound
\begin{align}\label{eq:Bellsteeringbound}
C_n = \max_{\{A_k\}}  \cu{ \lambda_{\rm max} \ro{\frac{1}{n}\sum_{k=1}^n
\hat{\sigma}_k^A{B_k}}},
\end{align}
where $\lambda_{\rm max}(\hat O)$ denotes the largest eigenvalue of $\hat O$.

An approach to constructing this family of steering inequalities is transforming from Bell inequalities. Bell states are shown to maximally violate analog CHSH inequality \cite{Cavalcanti2015,Supic2016,Gheorghiu2017}. For partial entangled two-qubit states, the authors in Ref. \cite{Shrotriya2021} constructed tilted steering inequalities from tilted-CHSH inequalities \cite{Acin2012}. In this paper,  we study the more general tilted steering inequalities construction from tilted-CHSH inequalities and study the robustness of one-sided self-testing based on analog steering inequalities. 
Furthermore, we consider to construct three measurement settings steering inequalities for general two-qubit states. 

\subsection{\label{sec:channel}  SDI certification and local extraction channel}
In this paper, we focus on one-sided self-testing two-qubit entangled state based on the steering inequalities. To this end, we first review the concept of self-testing.

Self-testing was originally known as a DI state verification,  where some observed statistics $p(a,b|x,y)$ from quantum devices can determine uniquely the underlying quantum state  and the measurements, up to a local isometry. As an example, the maximal violation of  CHSH inequality uniquely identifies  the maximally entangled  two-qubit state \cite{Mayers2004, McKague2012}. Usually, self- testing  relies on the observed  extremal correlations,
if the quantum systems that achieve the extremal correlations are unique up to local isometries, we say the extremal correlations $p(a,b|x,y)$ self- test the target system  $\{|\Bar{\psi}\rangle,\Bar{M}_{a|x},\Bar{N}_{b|y}\}$. Denoting the local isometry as $\Phi=\Phi_{AA'}\otimes\Phi_{BB'}$, self-testing can be formally defined as
\begin{equation}
    \begin{aligned}
    \Phi|\psi\rangle_{AB}{|00\rangle}_{A'B'}&=|\text{junk}\rangle_{AB}|\Bar{\psi}\rangle_{A'B'} \\
    \Phi M_{a|x}\otimes N_{b|y}|\psi\rangle_{AB}{|00\rangle}_{A'B'}&=|\text{junk}\rangle_{AB}\Bar{M}_{a|x}\otimes \Bar{N}_{b|y}|\Bar{\psi}\rangle_{A'B'}
    \end{aligned}
\end{equation}

Coming to the 1SDI scenario,
only the existence of an isometry $\Phi_B$ on Bob's side is required
\begin{equation}
\label{isometry}
    \begin{aligned}
    \Phi |\psi\rangle_{AB}{|0\rangle}_{B'}&=|\text{junk}\rangle_B\otimes|\Bar\psi\rangle_{AB'} \\
    \Phi M_{b|y}|\psi\rangle_{AB}{|0\rangle}_{B'}&=|\text{junk}\rangle_B\otimes \Bar{M}_{b|y}|\Bar\psi\rangle_{AB'}
    \end{aligned}
\end{equation}
where $M_{b|y}$ acts on $\mathcal{H}_{B}$; $\Bar{M}_{b|y}$ acts on $\mathcal{H}_{B'}$.

In addition to the above ideal definition of self-testing, it is essential to study the robustness of self-testing in the imperfect case when the obtained data deviate from the ideal value. There are two frameworks in the robustness analysis of self-testing. The first approach is based on the swap method by introducing an ancilla system.  The desired state can be swapped out of the real quantum system% based on local isometry equivalence,
then it could be calculated how far is it from the target state. One way to calculate this closeness is based on the analytic method involving mathematical inequalities techniques first proposed  in \cite{McKague2012}. The second one is the numerical method based on semidefinite programming combining NPA hierarchy \cite{Miguel2008}. Usually, the numerical method gives  much higher robustness.

The second approach is based on operator inequalities first introduced in Ref. \cite{Kaniewski2016}, which is now widely used in the robustness analysis of self-testing. For self-testing Bell state using CHSH inequality and self-testing GHZ state using Mermin inequality, the operator inequalities give nearly optimal bound. Robustness analysis of self-testing with operator inequalities can recur to  \textit{local extraction map}, which  hinges on the idea that local measurements can be used to virtually construct local extraction channel to extract the desired state from the real quantum system.
The local \textit{ extractability} of target $\psi_{AB}$ from  $\rho_{AB}$ is quantified
 \begin{equation}\label{eq:extractability}\Xi(\rho_{AB}\rightarrow\psi_{AB}):=\max\limits_{\Lambda_A,\Lambda_B}F((\Lambda_A\otimes\Lambda_B)(\rho_{AB}),\psi_{AB}),
 \end{equation}
where the maximum is taken over all possible local channels constructed with local measurements. For the 1SDI scenario,  Alice's side is trusted, thus the extraction channel in Alice's side is  $\Lambda_A= I_A$.
The lower bound of the fidelity between $\rho$ and the target state under the observed steering inequality can be defined as \textit{one-sided extractability}
 \begin{equation}\label{eq:fidelity}F(\rho_{AB},\psi_{AB}):=\inf\limits_{\rho_{AB}:S(\rho)\geq S_{obs}}\max\limits_{\Lambda_B}F(\Lambda_B(\rho_{AB}),\psi_{AB}),
 \end{equation}
 where $S(\cdot)$ is the  steering expression and   $S_{obs}$ is observed violation.  
To derive a linear bound of the fidelity about observed steering inequality violation,  real parameters $s$ and $\tau$ are required to be fixed such that $F\geq s\cdot S_{obs}+\tau$. This is equivalent to find $\Lambda_B$ (constructed by
Bob's local measurement operators  $M_y^b$) to make
\begin{equation}\label{eq:steering ineq}K\geq s S+\tau \mathbb{I}\end{equation}
where
$K:=(I_A\otimes\Lambda^+_B)(\psi_{AB})$ and $\Lambda^+$ refers to the dual channel of quantum channel $\Lambda$. By taking the trace with the input
state $\rho_{AB}$ on both sides of Eq. \eqref{eq:steering ineq}, one can get  $F\geq s\cdot S_{obs}+\tau$, in view of $\langle\Lambda_B^+(\psi_{AB}),\rho_{AB}\rangle 
=\langle \psi_{AB},\Lambda_B(\rho_{AB})\rangle$.

In the 1SDI scenario,  Bob's side is untrusted, thus Eq. \eqref{eq:steering ineq} is required to hold for Alice in two dimension and  Bob in arbitrary dimension. Since the measurements we considered in this paper is dichotomic, considering in qubit space will be sufficient in Bob's side.

\section{\label{sec:TCHSHsteering}One-sided self-testing based on 2-setting steering inequalities}
In device-independent scenario, general pure entangled two-qubit state
\begin{align}\label{eq:partial}
   \ket{\Phi}=\cos \theta \ket{00}+\sin \theta \ket{11}, 
\end{align}
has been proved to be self-tested \cite{Bamps2015,Coopmans2019} by the maximal violation of tilted-CHSH inequalities \cite{Acin2012}, which can be parametrized as
\begin{align}\label{eq:TCHSH}
   \Hat{I_\alpha}=\alpha A_0+ A_0B_0+ A_0B_1+A_1B_0-A_1B_1 \leq \alpha + 2,
\end{align}
where $\sin{2\theta}=\sqrt{\frac{4-\alpha^2}{4+\alpha^2}}$. The maximum quantum value is $\sqrt{8+2\alpha^2}$. The quantum measurements used to achieve the maximal quantum violation are: $\{\sigma_z;\sigma_x\}$ for Alcie, and $\{\cos\mu\sigma_z+\sin\mu\sigma_x\nonumber;\cos\mu\sigma_z-\sin\mu\sigma_x\}$ for Bob, where $\tan \mu=\sin 2\theta$ and $\sigma_{x,z}$ are Pauli $X,Z$ measurements. 

When $\alpha=0$, it corresponds to CHSH inequality and the state can be self-tested as Bell state. 
The self-testing criteria based on this tilted-CHSH inequalities is robust against noise. The best robustness bound to date can be found in \cite{Kaniewski2016,Coopmans2019}, in which the authors introduced the local extraction channel method.  However, as claimed in \cite{Kaniewski2016}, the theoretical optimal  upper bound is not achievable. Theoretically, the optimal bound is tied to the maximum classical violation which starts to achieve nontrivial fidelity. The nontrivial fidelity that demonstrates entanglement for the target state is $F>\cos^2 \theta$. They guessed that it might be related to the fact that the quantum value of the CHSH inequality does not reach its algebraic limit of 4. Here in 1SDI scenario, we will show that the theoretical optimal bound can be achieved.

To achieve 1SDI self-testing criteria, we will construct two types of 2-setting steering inequalities, which are based on above tilted-CHSH inequality by taking the measurements on Alice's side as trusted. 

\subsection{\label{subsec:TCHSHsteering}One-sided self-testing based on standard tilted-CHSH steering inequality}
Taking the measurements on Alice's side as trusted, the standard tilted-CHSH inequality in Eq. \eqref{eq:TCHSH}  can be transformed to the analog of tilted-CHSH steering inequality
\begin{align}\label{eq:TCHSHsteering}
   \Hat{S}_\alpha &=\alpha A_0+ A_0B_0+ A_0B_1+A_1B_0-A_1B_1 \nonumber\\
    &=\alpha {Z}+{Z(B_0+B_1)}+{X(B_0-B_1)} \nonumber \\
    & \leq \alpha + 2,
\end{align}
which maintains the maximum quantum violation $S_\alpha^{Q}=\sqrt{8+2\alpha^2}$ as in DI scenario. {We prove that partial entangled two-qubit states can be self-tested using this analog tilted-CHSH steering inequality in 1SDI manner. The proof is similar to DI self-testing using tilted-CHSH inequality, except that we can trust Alice's measurements now. The trustworthy of Alice's side can simplify the proof as an advantage. Another advantage is that theoretical optimal robustness bound can be obtained in 1SDI scenario with this steering inequality. By contrast, the optimal bound can not be achieved in DI self-testing with tilted-CHSH inequality. In the following, we will show both the analytical proof and the robustness analysis.} 
\paragraph{self-testing based on analog tilted-CHSH steering inequality}
%Here we provide a sketch of the proof for the self-testing statement based on titled-CHSH steering inequlity; the detailed proof is given in Appendix B 
%\yukun{The proof is straightforward  with sum of squares (SOS) decomposition tool which given in \cite {Bamps2015} for the family of Bell-like inequalities. }

\begin{figure}
	\begin{center}
		\includegraphics[width=8cm]{%1SSwap.png
		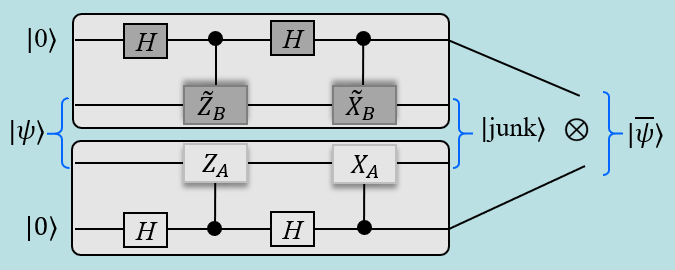}
		\caption{\label{fig:1sdiSwap}
			The SWAP isometry applied on Alice and Bob's side, where the operators $Z_A$ and $X_A$ are exactly the Pauli $Z,X$ operators.
		}
	\end{center}
\end{figure}

{We provide the simple proof here. Though Alice's side are trustworthy, as definition only the existence of isometry in Bob's side will efficient to determine uniquely the state and the measurements. However, for simplicity, we also introduce one isometry in Alice's side, which has been widely used in DI scenario, shown in Fig. \ref{fig:1sdiSwap}. As shown in bellow, with sum of
squares decomposition of positive semidefinite matrix \cite{Peyrl2008}, it's easy to find the algebraic relations that are necessarily satisfied by target quantum state and measurements to complete the proof.   } 

 After the isometry, the systems will be 
\begin{align}\label{eq:isometry}
   \Phi(\ket{\psi}) &=\frac{1}{4}[(I+Z_A)(I+\tilde{Z}_B)\ket{\psi}\ket{00} \nonumber\\
    &+X_A(I+Z_A)(I-\tilde{Z}_B)\ket{\psi}\ket{01} \nonumber \\
    &+\tilde{X}_B(I-Z_A)(I+\tilde{Z}_B)\ket{\psi}\ket{10}\nonumber \\
    &+X_A\tilde{X}_B(I-Z_A)(I-\tilde{Z}_B)\ket{\psi}\ket{11}]
\end{align}
To derive the underlying state $\ket{\psi}$ is equivalent to the target one, the algebraic relations between the operator acting on the state should be given. 
{
We notice that the analog tilted-CHSH steering inequality
$  \Hat{S}_\alpha$ have the maximum quantum value
 $ S_\alpha^{Q}$. 
This implies that the operator $\widehat{\mathcal{S}}_{\alpha}:=
S_\alpha^{Q}\mathbb{I}- \Hat{S}_\alpha
$
should be positive semidefinite (PSD) for all possible quantum states and measurement
operators in Bob's side. This can be proven by providing a set of operators $\{P_i\}$
which are polynomial functions of $A_x$ ($Z_A, X_A$) and $B_y$
such that $
\widehat{\mathcal{S}}_{\alpha}=\sum_i P^\dagger_i P_i
$,
holds for any set of measurement operators satisfying the
algebraic properties $A^2_x=\mathbb{I}$, $B^2_y=\mathbb{I}$. The decomposition form of  $\widehat{\mathcal{S}}_{\alpha}=\sum_i P^\dagger_i P_i$  is called a
\textit{sum of squares}(SOS). By SOS decomposition one can provide a direct certificate
that the upper quantum bound of $\Hat{S}_\alpha$is $S_\alpha^{Q}$ from its PSD, as well as some  relations between the projectors on the states, which will be used to give self-testing statement. This method was first introduced in  \cite{Bamps2015} for  the family of CHSH-liked Bell inequalities. Given SOS decompostions,  if one observes the maximal quantum violation of the steering inequality (CHSH-liked one) under state $\ket{\psi}$,  then each squared terms in  SOS decompositions acting on $\ket{\psi}$ should be  zero, i.e., $P_i\ket{\psi}=0$. Then useful relations for the measurements operators acting on underlying state  can be obtained from  these zero terms.}

{Similar to CHSH inequality scenario, two types of SOS decompositions for analog tilted-CHSH operator in Eq. \eqref{eq:TCHSHsteering} can be given. The first one is
\begin{align}\label{SOS1}
\widehat{\mathcal{S}}_{\alpha}
=&\frac{1}{2\mathcal{S}^Q_{\alpha}}\{
\widehat{\mathcal{S}}_{\alpha}^2+( \alpha X_A-S_0)^2\}
\end{align}
And the second one is 
\begin{align}\label{SOS2}
\widehat{\mathcal{S}}_{\alpha}
=&\frac{1}{2 \mathcal{S}^Q_{\alpha}}
\big\{(2Z_A-\mathcal{S}^Q_{\alpha}\frac{B_0+B_1}{2}+\frac{\alpha}{2}S_1)^2\nn\\
&+(2X_A-\mathcal{S}^Q_{\alpha}\frac{B_0-B_1}{2}+\frac{\alpha}{2}S_2)^2\big\}
\end{align}
where \begin{align}
S_0&= Z_A(B_0-B_1)+X_A(B_0+B_1),\nn\\
S_1&=Z_A(B_0+B_1)-X_A(B_0-B_1),\\
S_2&=Z_A(B_0-B_1)- X_A(B_0+B_1).\nn
\end{align}
Based on the maximal violation of analog tilted-CHSH inequality, the existence of the SOS decomposition for $\widehat{\mathcal{S}}_{\alpha}$  implies :
\begin{align}
    Z_A|\psi\rangle-\tilde{Z}_B|\psi\rangle=0,\label{relation1}\\ 
\sin(\theta)X_A(I+\tilde{Z}_B)|\psi\rangle-\cos(\theta)\tilde{X}_B(I-Z_A)|\psi\rangle=0 \label{relation2}
\end{align}
where $\tilde{Z}_B:=\frac{B_0+B_1}{2\cos \mu}$, and $\tilde{X}_B:=\frac{B_0-B_1}{2\sin \mu}$. 
Then with the algebraic relation of (\ref{relation1})-(\ref{relation2}) and the fact that $Z_AX_A=-X_AZ_A$, the equation in Eq. (\ref{eq:isometry}) can be rewritten to \[\Phi(\ket{\psi})=\ket{\text{junk}}[\cos\theta\ket{00}+\sin\theta\ket{11}]\]
where $\ket{\text{junk}}=\frac{1}{2\cos\theta}(I+Z_A)\ket{\psi}$. This means the underlying state  are unique to the target one up to local isometries, 
thus completes the self-testing statement.}
%\frac{1}{4}[2(I+Z_A)\ket{\psi}\ket{00} \nonumber\\
 %     &+2\frac{\sin\theta}{\cos\theta}(I+Z_A)\ket{\psi}\ket{11}]\nonumber\\
 %     &\frac{1}{4}[2(I+Z_A)\ket{\psi}\ket{00} \nonumber\\
  %    &+2\frac{\sin\theta}{\cos\theta}(I+Z_A)\ket{\psi}\ket{11}]\nonumber\\

\paragraph{self-testing robustness}

 %In general we may not get correlations that
%maximally violate the steering  inequality but give a violation that is only close to maximal. In this case, the robustness of our self-testing criteria should be provided. 

{ Here we mainly focus on the self-testing of quantum states. For the  self-testing of quantum measurements,  the  analysis can be related to quantum states according to Ref. \cite{Yang2014}. The procedure is similar, starting with $ \Phi M_B(\ket{\psi})$ instead of $ \Phi(\ket{\psi})$. In this case, the figure of merit should quantify how $ M_B\ket{\psi}$ is  close to the ideal measurements acting on the target state.}
% Here we mainly focus on the states self-testing, for the  self-testing of the measurements (whose analysis can resort to \cite{Yang2013}). The procedure is similar, instead of starting with $\Phi M_B(\ket{\psi})$ other than  $\Phi(\ket{\psi})$. In that case, the figure of merit should
% quantify how $ M_B\ket{\psi}$ is close to the ideal measurements acting on the target state.

As introduced in Sec. \ref{sec:channel}, to obtain the better self-testing robustness bound for the state, we should find the smallest value of $s$ while keeps $K-s\Hat{S}-\tau \mathbb{I}$ to be PSD. To this end, we first give the  spectral decomposition of   $\Hat{S}_\alpha$. Without loss of generality, we  write Bob's measurements  as
\begin{align}
B_r=\cos \mu \sigma_z +(-1)^r \sin \mu \sigma_x,
\end{align}
with $r \in \{0,1\}$ and $\mu \in [0,\pi/2]$.  Then the spectral decomposition of $\Hat{S}_\alpha$ is
\begin{align}\label{eq:CHSHsDecompose}
 \Hat{S}_{\alpha}=\sum \lambda_i |\psi_i\rangle \langle\psi_i|,\;\; i=1,2,3,4
\end{align}
with $\lambda_1^2+\lambda_2^2=8+2\alpha^2,\lambda_3=-\lambda_2,\lambda_4=-\lambda_1$.

  According to  different value ranges of $\mu$, the following two cases are discussed.

\textbf{Case 1:} $\cos2\mu\geq\frac{\alpha^2}{4}$ or equivalently $\mu \in [0, \arcsin \sqrt{\frac{4-\alpha^2}{8}}]$.

The eigenvalues of $\Hat{S}_{\alpha}$ have the form,
\[
\lambda_{1/2}=\pm\sqrt{\alpha^2+4\sin^2\mu}+2\cos\mu.
\]
The eigenvectors and the constraints for $\gamma$ and $\mu$ are
\[
\left\{
                \begin{array}{ll}                |\psi_1\rangle=\cos\gamma|00\rangle +\sin\gamma|11\rangle;\\
|\psi_2\rangle=\sin\gamma|00\rangle -\cos\gamma|11\rangle;\\
|\psi_3\rangle=\cos\gamma|01\rangle +\sin\gamma|10\rangle;\\

|\psi_4\rangle=-\sin\gamma|01\rangle +\cos\gamma|10\rangle.\\
                \lambda_1 \cos^2\gamma  +\lambda_2 \sin^2\gamma=\alpha+2\cos\mu\\
       \lambda_2 \cos^2\gamma  +\lambda_1 \sin^2\gamma=-\alpha+2\cos\mu\\           (\lambda_1-\lambda_2)\cos\gamma\sin\gamma=2\sin\mu \end{array}
              \right.
\]
with $\sin2\gamma=\frac{2\sin\mu}{\sqrt{\alpha^2+4\sin^2\mu}}$.

To obtain the optimal robustness bound, we consider the following local extraction channel on Bob's side: with the probability of $q_1$, he performs the identity operation $I$ on his qubit; with the probability of $q_2$, he performs $\sigma_z$ on his qubit. 
By this local extraction channel, the ideal state is transformed into
$K=q_1|\psi\rangle\langle\psi|+q_2\sigma_z|\psi\rangle\langle\psi|\sigma_z$. Denote $K-s\Hat{I_\alpha}-\tau \mathbb{I}$ as $G$. The PSD condition of $G$  requires that all the eigenvalues of it are non-negative, which points out

      \begin{equation}\label{qicondition}
       \frac{2\sin\mu s-C}{2\cos\theta\sin\theta}+\frac{1}{2}   \leq q_1\leq   \frac{2\sin\mu s+C}{2\cos\theta\sin\theta}+\frac{1}{2}, \end{equation}
      where
       \[\begin{array}{ll}
 C=&\sqrt{\cos^2\theta+(\beta_Q-(\alpha+2\cos\mu))s-1}\\ 
 &\cdot
\sqrt{\sin^2\theta+(\beta_Q-(-\alpha+2\cos\mu))s-1}\\
                \end{array}
\]
and  $\beta_Q=\sqrt{8+2\alpha^2}$. 

We can choose $q_1$ in the suitable range to saturate its upper bound, which makes $G$ to be PSD.
Meanwhile, we obtain the smallest value of $s$ as
\begin{equation}\label{s-value}
  s=\frac{1-\cos^2\theta}{\beta_{Q}-(2+\alpha)},
\end{equation}
and the corresponding value of $\tau$ is
\begin{equation}\label{tau-value}
 \tau=1-\sqrt{8+2\alpha^2}s, 
\end{equation}
which exactly equal to the theoretical optimal value. Thus we obtain the optimal robustness bound in the 1SDI scenario using the given extraction channel. Therefore,  it gives  the optimal robustness bound of self-testing based on analog tilted-CHSH steering inequality:
\begin{equation}\label{robuststeering}
\begin{split}
   F&=({\beta-\sqrt{8+2\alpha^2}})s+1\\
&=({\beta-\sqrt{8+2\alpha^2}})\frac{1-\frac{\sqrt{2}\alpha}{\sqrt{4-\alpha^2}}}{2\sqrt{8+2\alpha^2}-(4+2\alpha)}+1   
\end{split}
\end{equation}
 for observed violation $\beta$.

\textbf{Case 2:}   $0\leq\cos2\mu\leq\frac{\alpha^2}{4}$ or equivalently $\mu \in ( \arcsin \sqrt{\frac{4-\alpha^2}{8}} , \frac{\pi}{4}]$. 

The local extraction channel in this case is: Bob performs identity operation $I$  with  probability  $q_1$, and performs $\sigma_z$  with  probability  $q_2$.  Then the ideal state is transformed into
$K=q_1|\psi\rangle\langle\psi|+q_2\sigma_x|\psi\rangle\langle\psi|\sigma_x$. The PSD condition of $G:=K-s\Hat{I_\alpha}-\tau \mathbb{I}\geq 0$ gives $q_1=$
\begin{equation*}
\text{max}\Big\{0,\frac{4\sin^2\mu  \cdot s^2+(C_1s+\tau)(C_2s-\tau)}{(\beta_Q+2\sin 2\theta \sin\mu+\cos^2\theta C_2-\sin^2\theta C_1)s-1}\Big\}
\end{equation*}
where $\beta_Q=\sqrt{8+2\alpha^2}$. Meanwhile it gives
$s=\frac{1-\cos^2\theta}{\beta_{Q}-(2+\alpha)}$, $\tau=1-\sqrt{8+2\alpha^2}s$, which turn out to obtain the same robustness bound  as in Case 1. See Appendix \ref{appendixA} for the  details.

In conclusion, the theoretical linear optimal robustness bound can be obtained for self-testing of  two-qubit entangled states using the analog tilted-CHSH steering inequality. Different from self-testing in DI scenario, theoretical optimal robustness bound can be obtained using local extraction channel method. The reason might be that extraction channel is needed only on one side in steering scenario without coordination.

\textbf{\textit{Comparison  with DI and DD scenario}} To put our work in perspective, we
compare the certification in the 1SDI scenario with both DI and device-dependent (DD) scenario. 

In the DD scenario, the measurements on both sides are trusted and equal to the ideal measurements. In this case, we have
\begin{align}
 \Hat{I_\alpha} &=\alpha A_0+ A_0B_0+ A_0B_1+A_1B_0-A_1B_1\\
 &=\alpha{Z}+2\cos{\mu}{ZZ}+2\sin{\mu}{XX}    % \expval
\end{align}
where $\sin{2\theta}=\sqrt{\frac{4-\alpha^2}{4+\alpha^2}}$ and $\tan{\mu}=\sin{2\theta}$. It could be shown that
\begin{align}
\dyad{\Psi} \geq \frac{\Hat{I_\alpha}}{\sqrt{8+2\alpha^2}}. 
\end{align}

Thus in trusted measurement scenario, we have the lower bound of the fidelity
\begin{equation}
 F_{\text{DD}} \geq \frac{\beta}{\sqrt{8+2\alpha^2}}.  
\end{equation}
   
In the DI scenario, the authors in \cite{Coopmans2019} conjectured the lower bound of fidelity
\begin{equation}
    F_{\text{DI}} \geq s_{\alpha}\beta+\mu_{\alpha},
\end{equation}
with 
\begin{align}
s_{\alpha} &= \frac{ \big( \sqrt{ 8 + 2 \alpha^{2} } + 2 + \alpha \big) \big( 3 \sqrt{ 8 + 2 \alpha^{2} } - \sqrt{ 4 - \alpha^{2} } - \alpha \sqrt{ 2 } \big) }{4 ( 2 - \alpha )^2 \sqrt{ 8 + 2 \alpha^{2} }},  \\
 \mu_{\alpha} &= 1 - s_{\alpha} \cdot \sqrt{ 8 + 2\alpha^2 }. 
\end{align}

\begin{figure}	\includegraphics[height=9.5cm,width=9.5cm]{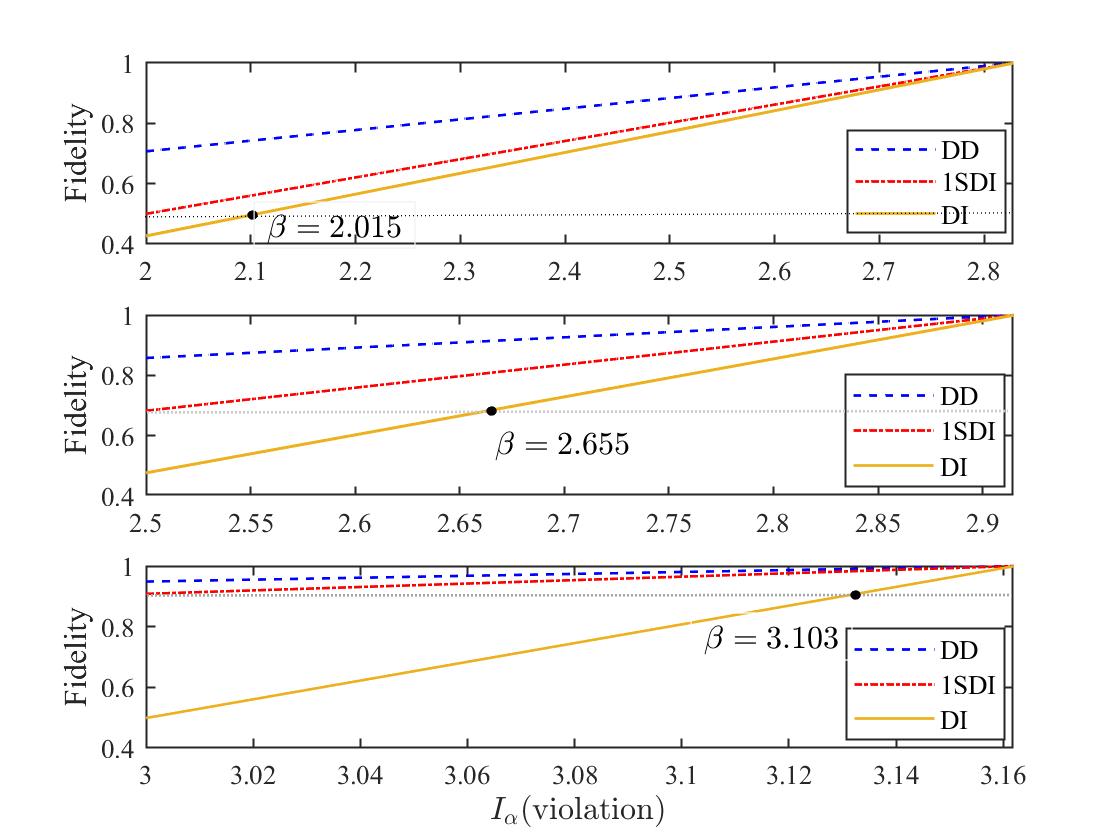}
	\caption{\label{comparefig}
			The comparison of robustness bound between DI (solid yellow  line), 1SDI (dotted-dashed red line), and DD (dotted blue line,device dependent) for different value of $\alpha$,which are $0,0.5,0.1$ from top  to bottom. 
		}
\end{figure}

Their comparison with SDI scenario is given in Fig. \ref{comparefig}. In the case of $\alpha=0$, it corresponds to CHSH inequality and the target state is singlet. Other two cases correspond to tilted-CHSH inequality and partially entangled two-qubit states. Obviously, it has $F_{\text{DD}}>F_{\text{1SDI}}>F_{\text{DI}}$ for all three cases. For $\alpha=0$, the nontrivial fidelity bound of singlet state  is $0.5$. The results  show that nontrivial fidelity bound can be obtained in DI scenario when the quantum value is larger than $2.105$, while for 1SDI and DD scenario the bound is  $2$ and $\sqrt{2} $ respectively. 
For $\alpha=0.5$, the nontrivial fidelity bound of target state  is $0.672$. The results  show that nontrivial fidelity bound can be obtained in DI scenario when the quantum value is larger than $2.655$, while for 1SDI and DD scenario the bound is   $2.5$ and $1.958 $ respectively. 
For $\alpha=1$, the nontrivial fidelity bound of target state  is $0.816$. The results  show that nontrivial fidelity bound can be obtained in DI scenario when the quantum value is larger than $3.103$, while for 1SDI and DD scenario the bound is   $3$ and $2.581 $ respectively. It is shown that with the increase of $\alpha$, especially for $\alpha=1$, the 1SDI self-testing bound is much better than DI scenario and more close to the DD scenario. Thus our method achieves significant improvement in the 1SDI certification of less entangled two-qubit states, which is comparable to the device-dependent scenario.
\subsection{One-sided self-testing based on general tilted-CHSH inequality}
In this section, we construct 2-setting steering inequalities from general
tilted-CHSH inequality\cite{Acin2012}
\begin{align}\label{eq:TCHSH2}
\mathcal{\hat{S}}_{\alpha,\beta}=\alpha A_0+\beta A_0B_0+\beta A_0B_1+A_1B_0-A_1B_1.
\end{align}
The maximal classical and quantum bounds are $\alpha+2(1+\beta)$ and $\sqrt{(4+\alpha^2)(1+\beta^2)}$, respectively.
The quantum bound can be achieved by pure two-qubit states \eqref{eq:partial} and corresponding measurements settings $\{\sigma_z;\sigma_x\}$ for Alcie, and $\{\cos\mu\sigma_z+\sin\mu\sigma_x\nonumber;\cos\mu\sigma_z-\sin\mu\sigma_x\}$ for Bob. with $\sin2\theta=\sqrt{\frac{4-\alpha^2\beta^2}{4+\alpha^2}}$ and $\tan\mu=\frac{\sin2\theta}{\beta}$.

 Taking the measurements on Alice's side as trusted, this Bell inequality can be transformed into
 \begin{align}
   \Hat{S}_{\alpha,\beta} 
    &=\alpha {Z}+\beta{Z(B_0+B_1)}+{X(B_0-B_1)} 
\end{align}
 which is a steering inequality. However we can also introduce two other measurements  to represent $B_0+B_1$ and $B_0-B_1$, thus rewrite the steering inequality as, 
 \begin{align}\label{eq:TCHSHsteering2_3}
   S_{\alpha,\beta}^{(1)}&=\alpha \expval{Z}+ \beta \expval{Z B_0}+  \expval{X B_1} \leq  \sqrt{1+(\alpha+\beta)^2}
\end{align}
with $\beta>0$. The maximal quantum violation is $\beta+\sqrt{1+\alpha^2}:=S_Q$. 

With this form of steering inequality, it allows us to   compare the construction with the one proposed in Ref. \cite{Shrotriya2021}, which changes the marginal term to Bob's side,
  \begin{align}\label{eq:TCHSHsteering2_2}
   S_{\alpha,\beta}^{(2)} &=\alpha \expval{B_0}+ \beta \expval{Z B_0}+  \expval{X B_1} \leq \alpha+ \sqrt{1+\beta^2},
\end{align}
with
$\beta^2=\alpha^2+1$, and keeps the quantum bound as Eq. \eqref{eq:TCHSHsteering2_3}. 
It should be remarked the constraints of  $\beta $ and $\alpha$ given in \cite{Shrotriya2021} can be  relaxed to  $\beta^2\geq\alpha^2+1$, which  we have proved in Appendix \ref{appendixD} with SOS decomposition related to the steering operators.  

{Both of these two steering inequalities of $S_{\alpha,\beta}^{(1)}$ and$ S_{\alpha,\beta}^{(2)}$ can be used to self-test pure partially entangled state  with $\sin(2\theta)=\frac{1}{\sqrt{1+\alpha^2}}$.The only difference between our construction and
the one in \cite{Shrotriya2021} is the exchanging role of Alice and Bob. 
The advantage of our construction will be shown later. Before that, we should give a proof about the maximum violation of both $S_{\alpha,\beta}^{(1)}$ and$ S_{\alpha,\beta}^{(2)}$ can be used to self-test pure partially entangled state. Though the proof for self-testing based on $S_{\alpha,\beta}^{2}$ has already been given in \eqref{eq:TCHSHsteering2_3}. However, a different proof is provided here which is based on the  SOS decomposition related to the steering inequality and the isometry given in Fig. \ref{fig:1sdiSwap}. The benefit with this proof is that  the constraints of  $\beta^2=\alpha^2+1$ can be relaxed, the details can been seen in Appendix \ref{appendixD}.}
%can be self-tested both when the steering inequality of $S_{\alpha,\beta}^{(1)}$ and$ S_{\alpha,\beta}^{(2)}$ are maximally violated. 

In the following we study the robustness of the self-testing based on these two steering inequalities.
%Thus the only difference between our construction and the one in \cite{Shrotriya2021} is the exchanging role of Alice and Bob in the first term of the inequality construction.
In Ref. \cite{Shrotriya2021}, the robustness of one-sided self-testing is studied only for maximally entangled states based on operator inequalities. For the case $\alpha=0$, when the violation of the steering inequality is $S=2-\epsilon$, the actual state is $24\sqrt{\epsilon}+\epsilon$ close to the target state, see also Ref. \cite{Supic2020}. More precisely, the relation between the fidelity and the steering inequality value is
  \begin{align}\label{eq:TCHSHsteeringFid2021}
F \geq 1-24\sqrt{2-S}-(2-S),
\end{align}
which is quite loose. Nontrivial fidelity bound $f>1/2$ can only be obtained when the violation is larger than $1.99957$, which makes the robustness analysis in the one-sided self-testing impractical. Here we have improved this bound to be
\begin{align}
  F \geq \frac{S-2}{4-2\sqrt{2}}+1,
\end{align}
which is the theoretical optimal linear bound. The local extraction channel to achieve this bound is constructed in Appendix \ref{appendixB}, and this channel  coincides with the extraction channel in the DI scenario introduced in Ref. \cite{Kaniewski2016}. However, the reason why this channel is used is not explained in Ref. \cite{Kaniewski2016}. %and other papers  \cite{Sekatski2018,Zhao2020}.
Here we point out the channel is the optimal local channel that local party can take. 

For the other case of $\alpha$, we give the robustness analysis of one-sided self-testing based on the numerical  method. The details are given in Appendix \ref{appendixC}.  The method works for general pure two-qubit states and the results show that the robustness bound is nearly linear.

\begin{figure}
	\begin{center}		\includegraphics[width=8.8cm]{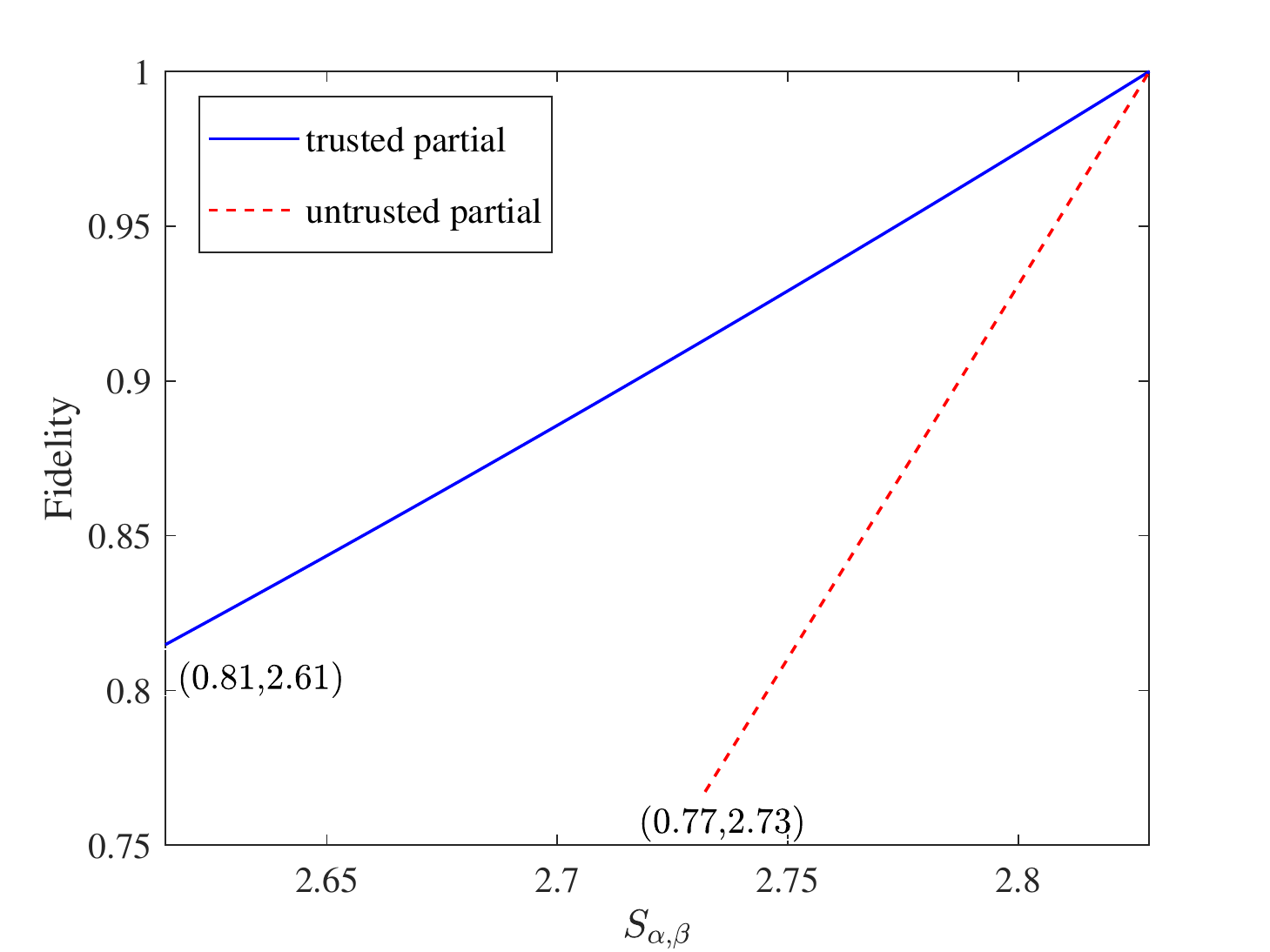}
		\caption{\label{fig:trust-untrust}
			 The comparison of the robustness bound  of self-testing based on 2-setting steering inequality $S_{\alpha,\beta}$ of Eq. (\ref{eq:TCHSHsteering2_3}) and Eq. (\ref{eq:TCHSHsteering2_2}), where  $\alpha=1$ and $\beta=\sqrt{2}$.   
		}
	\end{center}
\end{figure}

The comparison of the robustness bound  of self-testing based  of Eq. (\ref{eq:TCHSHsteering2_3}) and Eq. (\ref{eq:TCHSHsteering2_2}) are given in  Fig. \ref{fig:trust-untrust}, where we take $\alpha=1$ and $\beta=\sqrt{2}$ as an example. As shown, the one with trusted partial information gives a better robustness bound. The reason behind is that construction of steering inequality of Eq. \eqref{eq:TCHSHsteering2_3} shows smaller LHS bound compared with inequality of Eq. \eqref{eq:TCHSHsteering2_2}, however keeps the quantum maximum bound. Thus inequality of Eq. \eqref{eq:TCHSHsteering2_2} demonstrates an advantage for self-testing, it is more robust compared to using untrusted parties partial measurement expectation. Actually, in addition to the advantage in self-testing, the steering inequality constructed with trusted partial expectation can also have fewer constraints on variants $\alpha$ and $\beta$, thus could provide more  reasonable steering inequalities, see Appendix \ref{appendixD} for details.

\section{One-sided self-testing based on 3-setting steering inequalities}
So far the steering inequalities we considered are all of two measurement settings. In this section we introduce more measurements settings in constructing steering inequalities. Later it shows that adding more measurement settings can help to increase the robustness in one-sided self-testing. We construct a family of three setting steering inequalities
\begin{equation}\label{three-setting-I}
    I_{\alpha,\beta}\equiv \alpha\langle Z\rangle+\beta\langle ZB_0\rangle+\langle XB_1\rangle +\langle YB_2\rangle \leq\sqrt{2+(\alpha+\beta)^2}
\end{equation}
where $\beta\geq0$. These inequalities can be viewed as a generalization of analog tilted-CHSH steering inequalities in Eq. \eqref{eq:TCHSHsteering2_2}. A third measurement involving Pauli $Y$ measurement is added. Similar to the discussion in  two setting scenario, the partial expectation in the construction can also  be untrusted party Bob's measurement $B_0$. Thus $I_{\alpha,\beta}\equiv \alpha\langle B_0\rangle+\beta\langle ZB_0\rangle+\langle XB_1\rangle +\langle YB_2\rangle$ is constructed. {These two slightly different  inequalities have different LHS bound while keep the same quantum bound, for the details discussion and their proof for self-testing two-qubit partial entangled state please see Appendix  \ref{appendixD}}. 

Here we just consider the first case in the main text for simplicity and give its self-testing robustness bound. The LHS bound is the maximum violation  that  we can have, assuming Bob has a pre-existing state known to Alice, rather than half of an entangled state shared with Alice. Bob's system may derived from a classical systems, thus we can  denote his corresponding declared result by random variable $B_k\in \{-1,1\}$  for  $k=0,1$.    As shown in \cite{Saunders2010}, it is easy to see that 
\begin{equation}
 \begin{array}{ll}
   I_{\text{LHS}}=\max_{B_k}\lambda_{\text{max}}( I_{\alpha,\beta}),
\end{array}
\end{equation}
$\lambda_{\text{max}}(\hat{O})$denotes the largest eigenvalue of $\hat{O}$. Then the LHS bound of Eq. \eqref{three-setting-I} shows to be   $
\sqrt{2+(\alpha+\beta)^2}$.

 The maximum quantum bound is $\beta+\sqrt{4+\alpha^2}:=S_Q$. This can be verified by the fact that $S_Q\mathbb{I}- I_{\alpha,\beta}$ is PSD. More precisely,
 \begin{equation}
 \begin{array}{ll}S_Q\mathbb{I}- \hat{I}_{\alpha,\beta}&=\frac{\beta}{2}(\mathbb{I}-ZB_0)^2\\
& +\frac{\sqrt{\alpha^2+4}}{4}(\mathbb{I}-\frac{\alpha}{\sqrt{4+\alpha^2}}Z-\frac{2}{\sqrt{4+\alpha^2}}XB_1)^2\\
& +\frac{\sqrt{\alpha^2+4}}{4}(\mathbb{I}-\frac{\alpha}{\sqrt{4+\alpha^2}}Z-\frac{2}{\sqrt{4+\alpha^2}}YB_2)^2\end{array}
\end{equation} 
The quantum systems used to achieve the maximal quantum violation are, $B_0=Z,B_1=X,B_2=-Y$ and $\ket{\Phi}=\cos \theta \ket{00}+\sin \theta \ket{11}$ with $\sin2\theta=\frac{2}{\sqrt{4+\alpha^2}}$, which in turn can be self-tested when the maximum violation is reached up, { see Appendix  \ref{appendixD}}. % 

Here for simplicity, we just consider the case of $\alpha=0,\beta=1$. Assume Bob's measurements are untrusted, without loss of generality, they can be written as, $B_{0,1}=\cos{\mu}\sigma_z\pm \sin{\mu}\sigma_x$ and 
$B_2=\cos\mu_1\cos\mu_2\sigma_z+\cos\mu_1\sin\mu_2\sigma_x+\sin\mu_1\sigma_y$. Due to the asymmetric of $I_{\alpha,\beta}$ introduced by the form of $B_2$, the spectral decomposition of it is not easy, which leads to the difficulty in constructing local extraction channel making $G$ PSD.  We divide $G$ into two parts. If each part is PSD, then the whole matrix $G$ is PSD.
\begin{align}
\label{eq:operator}
G:=&K-( s (ZB_0+XB_1+YB_2)+\tau \mathbb{I})\nonumber\\
=&K_1- s (ZB_0+XB_1)-\tau_1 \mathbb{I}\\
&+K_2-sYB_2-\tau_2 \mathbb{I}\nonumber 
\end{align}
where $K_1+K_2=K$ denotes the two parts.

We consider the local extraction channel which ensures the part of $G_1:=K_1- s (ZB_0+XB_1)-\tau_1 \mathbb{I}$ and $G_2:=K_2-sYB_2-\tau_2 \mathbb{I} $ PSD simultaneously, see Appendix \ref{appendixF} for the details of the channel construction. The following robustness bound of self-testing in 3-setting steering scenario is obtained
\begin{align}\label{CHSHsFidelity}
  F \geq s{S_{obs}}+\tau \geq \frac{3}{12-4\sqrt{2}}{(S_{obs}-3)}+1.
  \end{align}
It should be noticed that, here we did not get the expected robustness bound of $F \geq \frac{(S_{obs}-3)}{2(3-\sqrt{3})}+1$.  
This may be because that the local extraction channel strategy we considered here is not optimal. It may be possible to find a better extraction strategy than here to obtain that bound. However, though the bound we give is optimal, it is still better than 2-setting analog-CHSH steering scenarios. 

\begin{figure}
	\begin{center}
		\includegraphics[width=8.8cm]{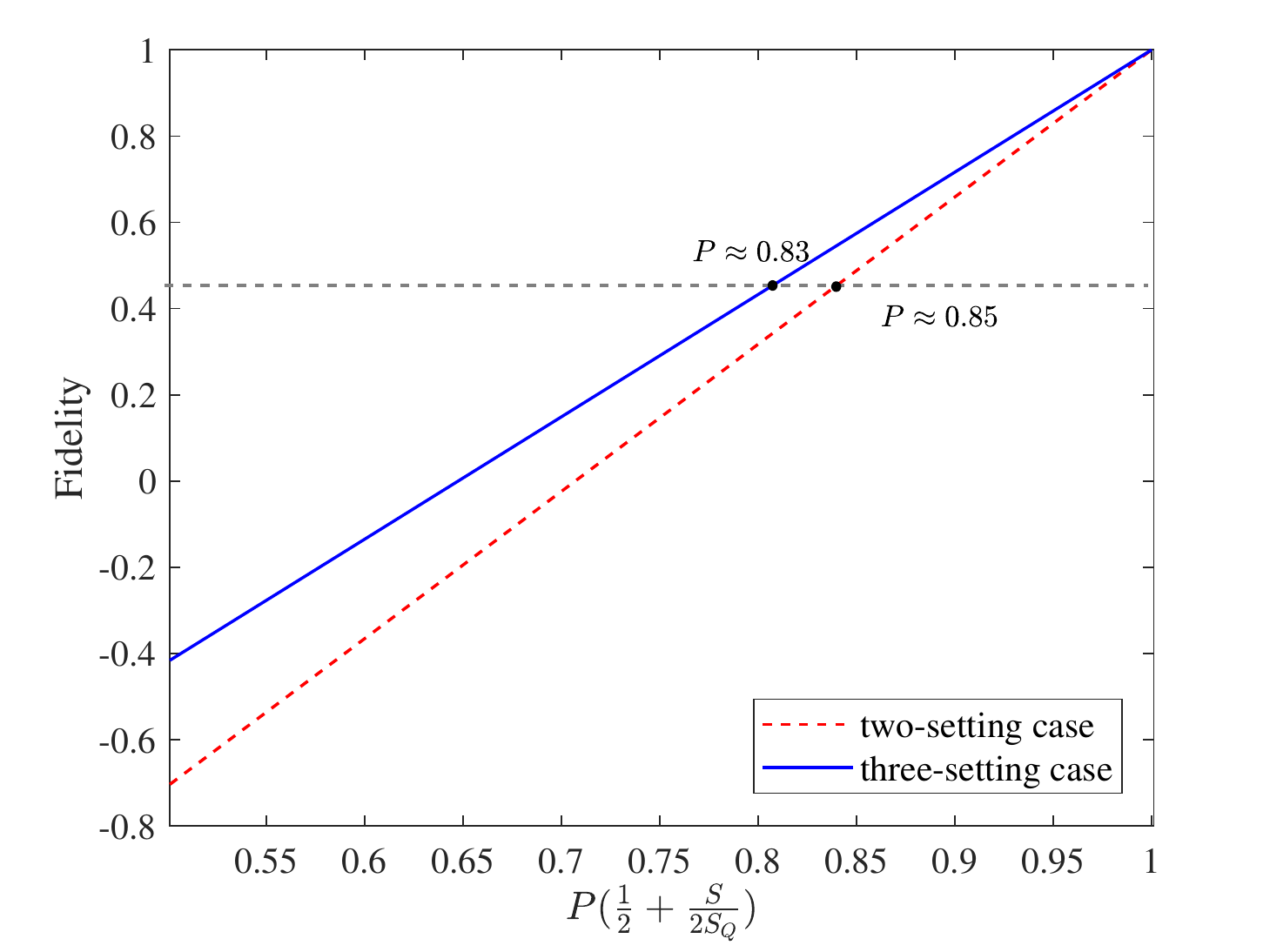}
		\caption{\label{fig:3better2}
			 Comparison of robustness bounds for one-sided self-testing of singlet based on  3-setting and 2-setting steering inequalities.   
		}
	\end{center}
\end{figure}

For a straightforward comparison between different inequalities, we transform the steering inequalities into the games characterized by the guessing probability which belongs to the same interval $[1/2,1]$.
In the case of $\alpha=0$, we have $P=\sum_{i=0,1} p(a=b|A_iB_i)=\frac{1}{2}+\frac{S}{2S_Q}$, which is the successful probability of the nonlocal game  guessing the other party's outcomes. 
For the other case, we can also find a nonlocal game, namely  the guessing score is related to the inequality in Eqs. \eqref{eq:TCHSHsteering2_3} and \eqref{three-setting-I}, respectively. See Appendix \ref{appendixE} for details.
We define the guessing probability as the probability for untrusted parties to successfully guess the trusted parties' outcomes,  which is also important for the sample efficiency analysis in next section. Based on the guessing probability, we can compare the robustness bound for for one-sided self-testing of singlet based on 3-setting and 2-setting steering inequalities. The result is shown in Fig. \ref{fig:3better2}, where the 3-setting steering inequality we constructed gives a better robustness bound. It is worthy to study whether steering inequalities with more measurement settings can be constructed and further improve the robustness of one-sided self-testing.

\section{\label{sec:sampleefficiency} Sample efficiency}
To construct a practical quantum verification protocol, it is crucial to study the sample efficiency \cite{Pallister2018,Zhu2019,Han2021,Dimic2022}.
Sample efficiency is used to study the performance of the self-testing criteria in the finite copy regime, in a way that  a fragment of the state copies is measured to warrant the rest states to be close to the target state.

 Consider a quantum device produces the states $\rho_1,\rho_2,\dots,\rho_N$ in $N$ runs. Our task is to verify  whether these states are sufficiently close to the target state $\ket{\Phi}\in \mathcal{H}$ on average.  Here the one-sided extractability is a natural choice for quantifying the closeness in one-sided self-testing scenario. 

 For the extraction channel method, we obtain linear relation between the extractability and the observed value of the steering inequalities
\begin{align}
    F\geq s\cdot S_{obs}+\tau.
\end{align}
Since $\tau=1-s\cdot S_Q$, we have
\begin{align}\label{eq:violation-fidelity}
  s\cdot (S_Q - S_{obs}) \geq 1-F.
\end{align}

The first step to construct the verification protocol is to view the steering inequalities as testing games. The details of transforming steering inequalities to testing games are shown in Appendix \ref{appendixE}.  Based on this, results on unmeasured copies can be guaranteed based on the  the measured copies.  Define $p$ as the guessing probability  of the game for a single state.  For the steering inequality in Eqs. \eqref{eq:TCHSHsteering2_3} and \eqref{three-setting-I}, when $\alpha=0$ which corresponds to the singlet state, the testing game is straightforward based on the outcomes of the same Pauli measurements. When $\alpha>0$ and corresponds to non-maximal entangled state, virtual testing games are constructed from the steering inequalities in Appendix \ref{appendixE}. For these testing games,  we have
\begin{align}\label{eq:guessingprob}
 p=\frac{1}{4}\sum_{i=0,1} p(a=b|A_iB_i)=\frac{1}{2}+\frac{S}{2S_Q}.   
\end{align}
This relation between the guessing probability and the violation of steering inequalities is essential for the study of sample efficiency. For the CHSH-analog steering inequality in Eq. \eqref{eq:TCHSHsteering}, we have $p=\frac{1}{4}\sum_{a\otimes b=ij} p(a,b|A_iB_i)=\frac{1}{2}+\frac{S}{4}$. This probability corresponds to the successful probability to win the game of  $a\otimes b=ij$ for Alice and Bob. For steering inequalities in Eq. \eqref{eq:TCHSHsteering} for $\alpha\neq0$ and  Eq. \eqref{eq:TCHSHsteering2_2}, we have not found corresponding  testing games. One may resort to other theories to study its performance in the finite regime, such as  \cite{Bancal2021}.

Define $\epsilon=1-F$ as the infidelity, combining Eq. \eqref{eq:violation-fidelity} and Eq. \eqref{eq:guessingprob}, we have
\begin{align}
  p \leq 1-\frac{\epsilon}{2s \cdot S_Q}.
\end{align}
Define $c=\frac{1}{2s \cdot S_Q}$, in general we have
\begin{align}
  p \leq 1-c \epsilon.
\end{align}

Now for these inequalities which corresponds to a testing game, we are ready to estimate the number of copies sufficient to exceed a certain bound on the average one-sided extractability. Suppose the states in the test are independently distributed, the goal is to guarantee that the average one-sided extractability of the states  $\rho_1,\rho_2,\dots,\rho_N$  is larger than $1-\epsilon$  with  significance level $\delta$ (confidence level $1-\delta$).   According to Ref. \cite{Dimic2022}, the scaling of sample efficiency depends on whether the quantum bound and algebraic bound coincide for the games between participants. When the quantum bound and algebraic bound coincide, the number of copies satisfies \begin{align}
    N \geq \frac{\ln \delta^{-1}}{\ln (1-c\epsilon)^{-1}}\approx \frac{\ln \delta^{-1}}{c\epsilon}.
\end{align}

 For all the steering inequalities we considered in this paper, the 2-setting inequality in Eq. \eqref{eq:TCHSHsteering2_3} and the 3-setting inequality in Eq. \eqref{three-setting-I} satisfy this condition. In that case, the maximal guessing probability $1$ can be obtained in the testing games according to the strategy given in the Appendix. Thus we obtain the approximately optimal sample efficiency  for one-sided self-testing of general two-qubit states in both 2-setting and 3-setting case, which is comparable to the number needed in quantum state verification.

For the CHSH-analog steering inequality in Eq. \eqref{eq:TCHSHsteering},
the quantum bound and algebraic bound are different. The number of copies needed satisfies
\begin{align}
    N=O(\frac{\ln{\delta^{-1}}}{c^2\epsilon^2}),
\end{align}
according to Ref. \cite{Dimic2022}.

In this section, we studied the sample efficiency for one-sided self-testing of two-qubit entangled states. Based on the steering inequalities we constructed, approximately optimal sample efficiency can be obtained in the SDI scenario, which is comparable to the device-dependent scenario. For the general DI scenario, the scaling of testing number is usually in quadratic form. Thus our strategies demonstrate a significant advantage over DI self-testing in sample efficiency.

\section{\label{sec:Conclustion} Conclusion}

In this paper, we studied the one-sided self-testing of general pure two-qubit states in the untrusted quantum network in which one party is not honest. The self-testing strategies are based on the violation of quantum steering inequalities. To achieve this goal, we firstly study two setting scenarios, where the steering inequalities can be constructed from   standard tilted-CHSH inequalities and its general form. Based on these steering inequalities, we studied the robustness of one-sided self-testing using both local extraction map method and numerical semi-definite-programming method.  Especially, the local extraction map method shows to provide the analytical and theoretical optimal linear bound. Our result also demonstrates an explicit approach to construct the local extraction channel. The comparison with device-independent scenario and device-dependent scenario shows clearly that the robustness of SDI certification lies in the middle.  The numerical method involving SDP and swap trick gives nearly linear robustness bound for general pure two-qubit states. To construct a practical certification protocol, we also clarified the sample efficiency of our 1SDI self-testing protocols. The results show that approximately optimal sample efficiency can be obtained based on the steering inequalities we constructed.  

Furthermore, we construct three measurement settings steering inequalities for general two-qubit states, which is not studied for partially entangled state before. It is shown that the robustness bound can be further improved by introducing the third measurement setting. It is worthy to study whether steering inequalities with more measurement settings can be constructed and further improve the robustness of one-sided self-testing. This question is also of close interest to the foundational studies on quantum steering. The improvement of robustness bound in our work can be applied to the certification of high dimensional quantum devices as building blocks. In the future, it would be potential to generalize our results to generic bipartite pure states, multipartite GHZ states, and other quantum states.  \\

\begin{acknowledgments}

This research is supported by National Nature Science Foundation of China (Grant No.62101600, No.61901218, and No.62201252), China  University  of  Petroleum Beijing (Grant No.ZX20210019), State Key Laboratory of Cryptography Science and Technology(Grant No.MMKFKT202109), and Natural Science Foundation of Jiangsu Province, China (Grant No.BK20190407).\\

\end{acknowledgments}
\appendix

\section{Local extraction channel method for self-testing based on analog tilted-CHSH inequality} \label{appendixA}

This section provides the robust bound of the self-testing based on analog tilted-CHSH inequality in Case 2.

\textbf{Case 2:}   $0\leq\cos2\mu\leq\frac{\alpha^2}{4}$ or equivalently $\mu \in ( \arcsin \sqrt{\frac{4-\alpha^2}{8}} , \frac{\pi}{4}]$.

In this case, the egivenvalues of the decomposition of $ \Hat{S}_{\alpha}=\sum \lambda_i |\psi_i\rangle \langle\psi_i|$ is, $\lambda_{1,2}=\sqrt{\alpha^2+4\sin^2\mu}\pm 2\cos\mu$.
The constraints between $\gamma$ and $\mu$ are,
\[
\left\{
                \begin{array}{ll}
                \lambda_1 \cos^2\gamma  -\lambda_2 \sin^2\gamma=\alpha+2\cos\mu;\\
                    \lambda_2 \cos^2\gamma  -\lambda_1 \sin^2\gamma=\alpha-2\cos\mu;\\
                 (\lambda_1+\lambda_2)\cos\gamma\sin\gamma=2\sin\mu
                \end{array}
              \right.
\]
Still $\sin2\gamma=\frac{2\sin\mu}{\sqrt{\alpha^2+4\sin^2\mu}}$.

The local extraction channel in this case is: Bob takes rotation operation $I$  with  probability  $q_1$, and takes $\sigma_z$  with  probability  $q_2$.  Then the ideal state is transformed into
$K=q_1|\psi\rangle\langle\psi|+q_2\sigma_x|\psi\rangle\langle\psi|\sigma_x$. The PSD requirement of $G:=K-s\Hat{I_\alpha}-\tau \mathbb{I}\geq 0$ gives \\
\begin{widetext}
\begin{eqnarray}
\begin{pmatrix}
q_1\cos^2(\theta)-C_1s-\tau & 0 & 0&q_1\frac{\sin2\theta}{2}-2\sin\mu s \\
0 & q_2\cos^2\theta-C_2s-\tau & q_2\frac{\sin2\theta}{2}-2\sin\mu s& 0\\
0 &q_2\frac{\sin2\theta}{2}-2\sin\mu s  & q_2\sin^2\theta+C_1s-\tau& 0\\
q_1\frac{\sin2\theta}{2}-2\sin\mu s  & 0 & 0&q_1\sin^2\theta+C_2s-\tau \\
\end{pmatrix}
\geq 0
\end{eqnarray}
\end{widetext}
where $C_1=\alpha+2\cos\mu$, and $C_2=\alpha-2\cos\mu$. The eigenvalues OF $G$ are,
\begin{align}
 &\lambda_{1,2}=\frac{G_{11}+G_{44}\pm\sqrt{(G_{11}-G_{44})^2+4G_{14}^2}}{2},\\
 &\lambda_{3,4}=\frac{G_{22}+G_{33}\pm\sqrt{(G_{22}-G_{33})^2+4G_{23}^2}}{2}.
\end{align}
which should be positive to make $G$ is PSD,
\begin{equation*}
\left\{
\begin{split}
&q_1\geq\frac{4\sin^2\mu  \cdot s^2+(C_1s+\tau)(C_2s-\tau)}{(\beta_Q+2\sin 2\theta \sin\mu+\cos^2\theta C_2-\sin^2\theta C_1)s-1}\\
&q_2 \geq \frac{4\sin^2\mu  \cdot s^2+(C_2s+\tau)(C_1s-\tau)}{(\beta_Q+2\sin2\theta\sin\mu+\cos^2\theta C_1-\sin^2\theta C_2)s-1}\\
\end{split}
\right.
\end{equation*}
where $\beta_Q=\sqrt{8+2\alpha^2}$.

We can also set
$s=\frac{1-\cos^2\theta}{\beta_{Q}-(2+\alpha)}$, and $\tau=1-\sqrt{8+2\alpha^2}s$, keeps  $q_1$ in above range. It gives the same bound as in Case 1. To this end, we take $ q_1$ to be the maximum between $0$ and the value which saturates the above  inequality in brace.  
%In conclusion, the robustness for self-testing in 1SDI scenario using analog tilted-CHSH steering inequality is the theoretical oprimal one.

\section{Local extraction channel method for self-testing based on reverse CHSH inequality} \label{appendixB}

For the analog CHSH steering operator $\Hat{S}= ZB_0+ XB_1$, it has the following spectral decomposition 
\begin{align}\label{eq:CHSHsDecompose}
 \Hat{S}=\sum \lambda_i |\psi_i\rangle \langle\psi_i| ,  
\end{align}
with $\lambda_1^2+\lambda_2^2=4,\lambda_3=-\lambda_2,\lambda_4=-\lambda_1$. Precisely, 
\begin{align}
&\lambda_1=\sqrt{2}(\cos \mu+\sin\mu),
\lambda_2=\sqrt{2}(\cos\mu-\sin\mu),
\end{align}
where Bob's measurements are written as $B_r=\cos \mu \sigma_z +(-1)^r\sin\mu \sigma_x$, with $r=0,1$.

In the case of  $\mu \in (0,\pi/4]$
, there has $\lambda_1,\lambda_2\geq 0$, and
\begin{equation}\label{CHSHdecom}
\left\{
                \begin{array}{ll}
|\psi_1\rangle=\frac{|00_B\rangle + |11_B\rangle}{\sqrt{2}};

|\psi_2\rangle=\frac{|00_B'\rangle + |11_B'\rangle}{\sqrt{2}}, \\

|\psi_3\rangle=\frac{|01_B'\rangle - |10_B'\rangle}{\sqrt{2}};

|\psi_4\rangle=\frac{|01_B\rangle - |10_B\rangle}{\sqrt{2}}, \; \text{where,}\\

0_B=\cos\frac{\pi}{8}|0\rangle+\sin\frac{\pi}{8}|1\rangle;

1_B=\sin\frac{\pi}{8}|0\rangle-\cos\frac{\pi}{8}|1\rangle\\

0_B'=\cos\frac{\pi}{8}|0\rangle-\sin\frac{\pi}{8}|1\rangle;

1_B'=\sin\frac{\pi}{8}|0\rangle+\cos\frac{\pi}{8}|1\rangle.
\end{array}
\right.
\end{equation}

We  consider the following local extraction channel: Bob takes rotation operation $R_{1}=I$ on his qubit with the probability of $q_1$, and takes $R_{2}=\sigma_z$ on his qubit with the probability of $q_2$ 
 , the ideal state is transformed into the mixture of Bell operator's eigenvectors
$|\psi\rangle:=q_1\ketbra{\psi_1}{\psi_1}+q_2\ketbra{\psi_2}{\psi_2}$.
In this case, $G:=K-s\Hat{S}-\tau \mathbb{I}$ is diagonal, the PSD requirement  gives,
\[
\left\{
                \begin{array}{ll}
                 q_1-s\lambda_1-\tau\geq 0 ;\\
                  q_2-s\lambda_2-\tau\geq 0 ;\\
                   \Tr(\rho)=p_1+p_2=1;\\
                \Tr(\rho \hat{S})=\lambda_1p_1+\lambda_2p_2=S;
                \end{array}
              \right.
\]
where we set $\tau=1-2s$.

By simplifying, we have
 $s\lambda_1-2s+1   \leq q_2\leq -s\lambda_2+2s$
which gives us $ s\geq\frac{1}{4-(\lambda_1+\lambda_2)} \geq\frac{1}{4-2\sqrt{2}}$.  It gives out the following robustness bound of self-testing  via steering inequality:
\begin{align}\label{CHSHsFidelity}
  F \geq s{S}+\tau \geq \frac{S-2}{4-2\sqrt{2}}+1. 
\end{align}
 Besides, we get the constaints on the rotation probability
 \begin{align}\label{CHSHsp1}
  (1+\sqrt{2})(\cos \mu+\sin \mu +1) \leq q_1 \leq 1. 
\end{align}

 For the case of $\mu \in (\frac{\pi}{4},\frac{\pi}{2})$,  the local extraction channel  are considered as: Bob takes rotation  $R_{1}=I$ with the probability of $q_1$, and takes $R_{2}=\sigma_x$ with the probability of $q_2$. It gives the same robustness bound.

Above,  we get the optimal linear bound and  nontrivial fidelity can be obtained as long as the steering inequality is violated. But, as shown in Ref. \cite{Kaniewski2016}  that nontrivial fidelity bound could not be obtained at inequality violation at $2$, with this local extraction channel.  The reason might be that to define the appropriate extraction channel the two local sites  need coordinating. In the DI scenario, both sides are not trusted. The decomposition of Bell operator is related both to Alice and Bob's local measurements directions. 

Once Alice and Bob could inform each other what measurement directions they choose (do classical communication), it is possible for them to define the appropriate local rotation channel which could rotate the idea states  to be the  eigenvetors of Bell operator with positive positive eigenvalues. It  could make $G:=K-s\Hat{I}-\tau \mathbb{I}$ to be PSD.  In this case, the  optimal $s$ and $t$ is easy to find to be the optimal one. However, allowing communication is no usual sense of device independent.  Thus in DI scenario, when there need coordination, the non-trivial fidelity could not be reached up. 

\section{Numerical results utilising the SWAP isometry} \label{appendixC}

In this section, we consider the numerical method based on SDP to show the robustness of the self-testing via steering inequality, which has been widely  used in DI frameworks \cite{Yang2014,Wang2016}. A detailed robustness analysis is given for 3-setting steering inequalities. For 2-setting scenarios, only need to remove the third measurement in the code.

\begin{figure}
	\begin{center}
		\includegraphics[width=8cm]{%1SSwap.png
    		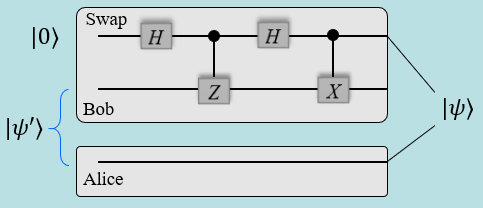}
		\caption{\label{fig:onesideSwap}
			The one-side SWAP isometry applied on Bob's side.   
		}
	\end{center}
\end{figure}

The target sate is $\ket{\psi}=\cos\theta\ket{00}+\sin\theta\ket{11}$. And  Bob's measurements  can be written as,
$B_0=2E_{0|0}-I$, $B_1=2E_{0|1}-I$ and $B_2=2E_{0|2}-I$, where $B_0^2=B_1^2=B_2^2$. After applying the isometry given in Fig. \ref{fig:onesideSwap} to the physical 
state $\ket{\psi'}$, we obtain the state
\begin{align}
    \ket{\psi'}=E_{0|0}\ket{\psi}\ket{0}_{A'}+XE_{1|0}\ket{\psi}\ket{1}_{A'}
\end{align}
We trace the desired system out
\begin{align}
    \rho_{\text{swap}}=\tr_A(\ket{\psi'}\bra{\psi'}).
\end{align} 

Utilising the SWAP isometry on Bob's side, the fidelity can be bounded as:
\begin{align}
f=&\bra{\psi}\rho_{\text{swap}}\ket{\psi} \nonumber\\
=&\cos^2\theta\bra{0} \tr_A(E_{0|0}\rho_{AB})\ket{0}
+\sin^2\theta \bra{1}\tr_A (E_{1|0}\rho_{AB})\ket{1}\nonumber \\
+&\frac{\sin2\theta}{2}[ \bra{0} \tr_A(E_{1|0}X E_{0|0}\rho_{AB})\ket{1}
+\bra{1}\tr_A(E_{0|0}X E_{1|0}\rho_{AB})\ket{0}]\nonumber \\
=&\cos^2\theta\bra{0} \tr_A(E_{0|0}\rho_{AB})\ket{0}+\sin^2\theta\bra{1}(\rho_B-\tr_A E_{0|0})\ket{1} \nonumber \\
&+\sin2\theta[\bra{0}\tr_A(E_{0|1}E_{0|0}-E_{0|0}E_{0|1}E_{0|0})\ket{1}\nonumber \\
&+\bra{1} \tr_A(E_{0|0}E_{0|1}-E_{0|0}E_{0|1}E_{0|0}\rho_{AB})\ket{0}]\nonumber \\
=&\cos^2\theta \bra{0} \sigma_{0|0}\ket{0}\nonumber+\sin^2\theta \bra{1}(\rho_B-\sigma_{0|0})\ket{1}\nonumber \\
&+\sin2\theta [\bra{0} (\sigma_{0|1,0|0}-\sigma_{0|0,0|1,0|0})\ket{1}\nonumber\\
&+\bra{1}\sigma_{0|0,0|1}-\sigma_{0|0,0|1,0|0}\ket{0}]
\end{align}

The goal is now to give a lower bound to $f$. % We will use the violation of 3-setting steering inequality to impose these constraints. For 2-setting scenarios, it just needs to remove the third measurement in the code.
The numerical method of minimizing the fidelity for given steering inequality value is given by the following SDP:
\begin{align}
\label{eq:sdp}
\textrm{minimize }&f:=\Tr(M\Gamma)\\ \nonumber
\textrm{subject to: }&\Gamma\geq 0,\\ \nonumber
&I_{\alpha,\beta} =Q,
\end{align}
where
$M$ is matrix $\text{zeros}(14,14)$, with
$M_{2,2} = \sin^2\theta$;
$M_{2,9} =M_{9,2} = \sin2\theta$;
$M_{3,3} = \cos^2\theta$;
$M_{4,4} = -\sin^2\theta$ and
$M_{9,10} =M_{10,9} = -\sin2\theta$.
% $I_{\alpha,\beta}=\alpha\langle Z\rangle+\beta\langle ZB_0\rangle+\langle XB_1\rangle +\langle YB_2\rangle=\Tr[(\alpha-\beta)Z\rho_C-(X+Y)\rho_C+2\beta Z\sigma_{0|0}+2X\sigma_{0|1}+2Y\sigma_{0|2}]$, or  $I_{\alpha,\beta}=\alpha\langle B_0\rangle+\beta\langle ZB_0\rangle+\langle XB_1\rangle +\langle YB_2\rangle=\Tr[-(\alpha+\beta Z+X+Y)\rho_C+(2\alpha I+2\beta Z)\sigma_{0|0}+2X\sigma_{0|1}+2Y\sigma_{0|2}]$. 
 
\begin{widetext}
\begin{align}
\Gamma&=\left( \begin{array}{ccccccc}
\rho_{C} & \sigma_{0|0} & \sigma_{0|1} &\sigma_{0|2} & \sigma_{0|1,0|0}&\sigma_{0|2,0|0}&\sigma_{0|2,0|1}\\
\sigma_{0|0} & \sigma_{0|0} & \sigma_{0|0,0|1}  &\sigma_{0|0,0|2} & \sigma_{0|0,0|1,0|0}& \sigma_{0|0,0|2,0|0}& \sigma_{0|0,0|2,0|1}\\
\sigma_{0|1} & \sigma_{0|1,0|0} & \sigma_{0|1} & \sigma_{0|1,0|2}& \sigma_{0|1,0|0}& \sigma_{0|1,0|2,0|0}& \sigma_{0|1,0|2,0|1}\\
\sigma_{0|2} & \sigma_{0|2,0|0} & \sigma_{0|2,0|1} & \sigma_{0|2}& \sigma_{0|2,0|1,0|0}& \sigma_{0|2,0|0}& \sigma_{0|2,0|1}\\
\sigma_{0|0,0|1} & \sigma_{0|0,0|1,0|0} & \sigma_{0|0,0|1} & \sigma_{0|0,0|1,0|2}& \sigma_{0|0,0|1,0|0}& \sigma_{0|0,0|1,0|2,0|0}& \sigma_{0|0,0|1,0|2,0|1}\\
\sigma_{0|0,0|2} & \sigma_{0|0,0|2,0|0} & \sigma_{0|0,0|2,0|1} & \sigma_{0|0,0|2}& \sigma_{0|0,0|2,0|1,0|0}& \sigma_{0|0,0|2,0|0}& \sigma_{0|0,0|2,0|1}\\
\sigma_{0|1,0|2} & \sigma_{0|1,0|2,0|0} & \sigma_{0|1,0|2,0|1} & \sigma_{0|1,0|2}& \sigma_{0|1,0|2,0|1,0|0}& \sigma_{0|1,0|2,0|0}& \sigma_{0|1,0|2,0|1}\\
\end{array} \right)
\end{align}

$I_{\alpha,\beta}=\alpha\langle Z\rangle+\beta\langle ZB_0\rangle+\langle XB_1\rangle +\langle YB_2\rangle=\Tr[(\alpha-\beta)Z\rho_C-(X+Y)\rho_C+2\beta Z\sigma_{0|0}+2X\sigma_{0|1}+2Y\sigma_{0|2}]$, \; or

$I_{\alpha,\beta}=\alpha\langle B_0\rangle+\beta\langle ZB_0\rangle+\langle XB_1\rangle +\langle YB_2\rangle=\Tr[-(\alpha+\beta Z+X+Y)\rho_C+(2\alpha I+2\beta Z)\sigma_{0|0}+2X\sigma_{0|1}+2Y\sigma_{0|2}]$. 
\end{widetext}

We constrain $\Gamma$ in the optimization to be positive semi-definite and note that each sub-matrix of $\Gamma$ corresponding to something like an element of an assemblage is a valid quantum object. It actually turns out that all assemblages that satisfy no-signalling can be realized in quantum theory \cite{Hughston}. Discussion of this point is beyond the scope of this paper as all we wish to do is give a lower bound on the value of $G$ therefore just imposing $\Gamma\geq 0$ gives such  bound. 
Based on the SDP of Eq. (\ref{eq:sdp}), we showed several robustness bound  of self-testing based on 3-setting steering inequality for $\alpha=1,2$ and $\beta=1,2,10$. 
\begin{figure}
	\begin{center}
		\includegraphics[width=9cm]{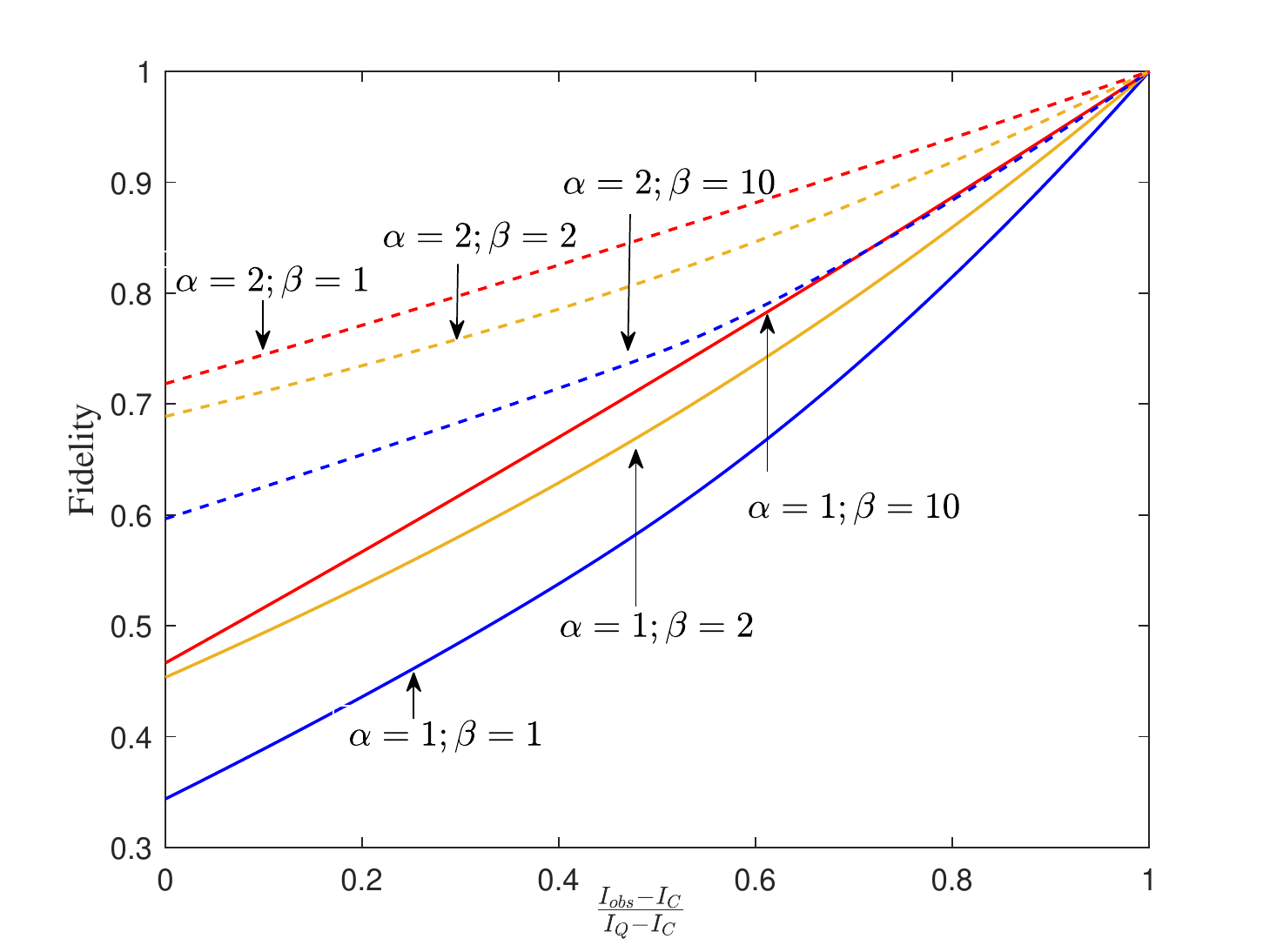}
		\caption{\label{fig:threesdp}
			 Robustness bound  of self-testing based on 3-setting steering inequality for $\alpha=1,2$ and $\beta=1,2,10$.   
		}
	\end{center}
\end{figure}

\section{Analysis of different type of 2-setting and 3-setting  steering inequalities}\label{appendixD}
{Here we study the maximal quantum violation of the steering inequalities 
 involved in the main text and provide that the maximal violation of these inequalities can be used to self-testing.}

For  2-setting steering inequality
  \begin{align}\label{eq:TCHSHsteering2_2appendix}
   S_{\alpha,\beta}^2 &=\alpha \expval{B_0}+ \beta \expval{Z B_0}+  \expval{X B_1} \leq \alpha+ \sqrt{1+\beta^2}
\end{align}
{The maximum quantum bound is $\beta+\sqrt{1+\alpha^2}:=S_Q$. This can be confirmed by showing $S_Q\mathbb{I}- \hat{S}_{\alpha,\beta}^{(2)}\geq0$ to be true for all the possible underlying state and the measurements.  To do so, we provide the following SOS decompositons of $S_Q\mathbb{I}- \hat{S}_{\alpha,\beta}^{(2)}$ to illustrate its PSD}.

{The first SOS decomposition is,
\begin{align} \label{sos1}
&S_Q\mathbb{I}- \hat{ S}_{\alpha,\beta}^{(2)}\nonumber\\
&=\alpha_1^2(\mathbb{I}-cB_0-sX_AB_1)^2+\alpha_2^2(Z_A-B_0)^2\nonumber\\
  &+\alpha_3^2(-cB_1+sX_AB_0+Z_AB_1)^2\nonumber\\
&+\alpha_4^2(S_Q\mathbb{I}-\hat{ S}_{\alpha,\beta}^2)^2
\end{align}
where $c=\frac{\alpha}{\sqrt{1+\alpha^2}}$,$s=\frac{1}{\sqrt{1+\alpha^2}}$,$\alpha_4^2=\frac{1}{4\beta}$,$\alpha_3^2=\frac{\beta\sqrt{1+\alpha^2}}{1}\alpha_4^2=\frac{\sqrt{1+\alpha^2}}{4}$,$\alpha_1^2=(\frac{\beta\sqrt{1+\alpha^2}}{1}-\frac{1+\alpha^2}{1})\alpha_4^2$, and $\alpha_2^2=\frac{\beta-\sqrt{1+\alpha^2}}{4}$.
%And the second one as,
%\begin{align} \label{sos2}
%&S_Q\mathbb{I}- \hat{ S}_{\alpha,\beta}^2\nonumber\\
%&=\alpha_1^2(\mathbb{I}-cB_0-sX_AB_1)^2+\alpha_2^2(Z_A-B_0)^2\nonumber\\
 % &+\alpha_3^2((1+s^2)B_0-2Z_A+cZ_AB_0-csX_AB_1)^2\nonumber\\
 % &+\alpha_4^2(-(1+s^2)B_1+2sX_A+cZ_AB_1-csX_AB_0))^2
%\end{align}
%where $\alpha_1$ and $\alpha_2$ are the same as the first SOS decomposition, and $\alpha_3^2=\alpha_4^2=\frac{S_Q}{8\beta s(1+s^2)}$.
And the second one is,
\begin{align} \label{sos2}
&S_Q\mathbb{I}- \hat{ S}_{\alpha,\beta}^{(2)}\nonumber\\
&=\alpha_1^2(\mathbb{I}-cB_0-sX_AB_1)^2+\alpha_2^2(Z_A-B_0)^2\nonumber\\  &+\alpha_3^2((\Delta+s^2)B_0-(\Delta+1)Z_A+cZ_AB_0-csX_AB_1)^2\nonumber\\
  &+\alpha_4^2(-(\Delta+s^2)B_1+s(\Delta+1)X_A+\Delta cZ_AB_1-csX_AB_0))^2
\end{align}
where $\alpha_1$ and $\alpha_2$ are the same as the first SOS decomposition, and $\alpha_3^2=\Delta \alpha_4^2$, $\alpha_4^2=\frac{S_Q}{4s\beta (\Delta^2+s^2)(\Delta^2+1)}$, $\Delta=\frac{\beta}{\sqrt{1+\alpha^2}}$.}

{It is easy to verify that the left part of Eq. (\ref{sos1})-(\ref{sos2})  are equal to the right SOS forms. In addition, to make the SOS decompositions are positive semidefinite, there should have $\alpha_i\geq 0$, thus has $\beta\geq\sqrt{1+\alpha^2}$. Apparently, $S_Q$ is the upper bound of the steering inequality ${ S}_{\alpha,\beta}^2$ under this constraint, although we don't know whether quantum can reach up the bound or not. Provided that  $B_0=Z,B_1=X$ and $\ket{\Phi}=\cos \theta \ket{00}+\sin \theta \ket{11}$ with $\sin2\theta=\frac{1}{\sqrt{1+\alpha^2}}$  can make ${ S}_{\alpha,\beta}^2$ achieves  $S_Q$, we conclude $S_Q$ is the maximum quantum violation.}

{Next, we show the maximal violation of this steering inequality will self-test the partial entangled state. The local isometry used to determine the equivalence of the states is the same as  the main text, but with $\tilde{Z}_B=B_0$ and $\tilde{X}_B=B_1$.  As shown in the main text, the relations required to show this isometry works are 
\begin{align}
    Z_A|\psi\rangle-B_0|\psi\rangle=0,\label{AP-relation1}\\
\sin\theta X_A(I+B_0)|\psi\rangle-\cos\theta B_1(I-Z_A)|\psi\rangle=0 \label{AP-relation2}
\end{align}
To obtain this relations, we let each side of  Eq. (\ref{sos1})-(\ref{sos2}) to take action on $\ket{\psi}$, which state was supposed to reach up the maximum violation of the steering inequality. Then seven  terms of $P_i\ket{\psi}=0$ will be obtained, among them the second squared term in  Eq. (\ref{sos1}) gives Eq.(\ref{AP-relation1}), meanwhile the linear combination of the third squared term in Eq.(\ref{sos1}) and the forth squared term in Eq. (\ref{sos2}) leads to Eq. (\ref{AP-relation1}).  Then similar to the proof for the analog of tilted-CHSH steering inequality given in the main text,  by  the  isometry given in Fig. \ref{fig:1sdiSwap},  we complete the self-testing statement via 2-setting steering inequality $S_{\alpha,\beta}^2$. 
%\begin{align}\label{reeq:isometry}
%   \Phi(\ket{\psi}) &=\frac{1}{4}[(I+Z_A)(I+\tilde{Z}_B)\ket{\psi}\ket{00} \nonumber\\
%    &+X_A(I-Z_A)(I+\tilde{Z}_B)\ket{\psi}\ket{01} \nonumber \\
 %   &+\tilde{X}_B(I-Z_A)(I+\tilde{Z}_B)\ket{\psi}\ket{10}\nonumber \\
 %   &+X_A\tilde{X}_B(I-Z_A)(I-\tilde{Z}_B)\ket{\psi}\ket{11}\nonumber \\
 %   &=\frac{1}{4}[2(I+Z_A)\ket{\psi}\ket{00} \nonumber\\
%      &+2\frac{\sin\theta}{\cos\theta}(I+Z_A)\ket{\psi}\ket{11}]\nonumber\\
 %     &=\ket{\text{junk}}[\sin\theta\ket{00}+\cos(\theta)\ket{11}].
%\end{align}
 }

For the 2-setting steering inequality
\begin{align}
   S_{\alpha,\beta}^{(1)} &=\alpha \expval{Z}+ \beta \expval{Z B_0}+  \expval{X B_1} \leq  \sqrt{1+(\alpha+\beta)^2}
\end{align}
It keeps the same maximal quantum violation as Eq. \eqref{eq:TCHSHsteering2_2appendix}. {For this steering inequality, three different types of SOS decompositins related to  $S_Q\mathbb{I}- \hat{S}_{\alpha,\beta}^{(1)}$ can be given, the first one is,
\begin{equation}\label{sos_z1}
\frac{\beta}{2}(\mathbb{I}-Z_AB_0)^2+\frac{\sqrt{\alpha^2+1}}{2}(\mathbb{I}-cZ_A-sX_AB_1)^2
\end{equation} 
the second one is
\begin{equation}\label{sos_z2}
\frac{1}{2S_Q}(-cX_A+sZ_AB_1+X_AB_0)^2+\frac{\beta\sqrt{\alpha^2+1}}{2S_Q}(S_Q\mathbb{I}- \hat{S}_{\alpha,\beta}^{(1)})^2
\end{equation}
and the third one is
\begin{align}\label{sos_z3}
&\alpha_1^2((\Delta+s^2)Z_A-(\Delta+1)B_0+cZ_AB_0-csX_AB_1)^2\nonumber\\
  &+\alpha_2^2(-(\Delta+s^2)X_A+s(\Delta+1)B_1+\Delta cX_AB_0-csZ_AB_1))^2
\end{align}
where $\alpha_1^2=\Delta \alpha_2^2$, $\alpha_2^2=\frac{(1+\alpha^2)^2}{2(\beta^2\sqrt{1+\alpha^2})+\beta(1+\alpha^2)+S_Q}$, $\Delta=\frac{\beta}{\sqrt{1+\alpha^2}}$.
The PSD requirements only require $\beta>0$.
And with each squared terms in Eq. (\ref{sos_z1})-(\ref{sos_z3}) acting on $\ket{\psi}$ is zero, it also can lead to the relations our self-testing proofs heavily relied on, namely Eq. (\ref{AP-relation1})-(\ref{AP-relation2})(the first term in Eq. \ref{sos_z1} leads to Eq. (\ref{AP-relation1}), the first term in  Eq. (\ref{sos_z1}) and the second term in Eq. (\ref{sos_z3}) lead to Eq. (\ref{AP-relation2})). Then we can complete the proof of self-testing based on $ S_{\alpha,\beta}^{(1)}$.}

For the 3-settings scenario, the partial part expectation can be changed into the untrusted part's measurement. Thus there are two 3-setting steering inequality, the one in the main text,
\begin{equation}\label{rthree-setting-I}
    I^{(1)}_{\alpha,\beta}\equiv \alpha\langle Z\rangle+\beta\langle ZB_0\rangle+\langle XB_1\rangle +\langle YB_2\rangle \leq\sqrt{2+(\alpha+\beta)^2}
\end{equation}
and the one,
\begin{equation}
    I^{(2)}_{\alpha,\beta}\equiv \alpha\langle B_0\rangle+\beta\langle ZB_0\rangle+\langle XB_1\rangle +\langle YB_2\rangle \leq \alpha+\sqrt{2+\beta^2},
\end{equation}
 The advantage of this change is its LHS bound is lower than using  Alice's $Z$ measurement in 3-setting inequality, while the quantum bound is maintained. It extends the gap between LHS bound and steering bound, which is a benefit for the practical experiment.  Denoting Bob’s corresponding declared result 
by the random variable $B_k\in \{-1,1\}$  for  $k=0,1$, it is easy to obtain the LHS bound 
$\alpha+\sqrt{2+\beta^2}$.

{The quantum bound of the both 3-setting steering inequality is the same,  $\beta+\sqrt{4+\alpha^2}$. However, an extra condition should be satisfied for $I^{(2)}_{\alpha,\beta}$, that is $\beta\geq \sqrt{4+\alpha^2}$. For $I^{(2)}_{\alpha,\beta}$ it only requires $\beta\geq0$. This can be obtained from the following SOS, the first one is, 
\begin{align}\label{3-settingSOS1}
&(\beta+\sqrt{4+\alpha^2})\mathbb{I}- \hat{I}^{(2)}_{\alpha,\beta}\nonumber\\
&=\alpha_1^2(\mathbb{I}-cB_0-sX_AB_1)^2+\alpha_2^2(Z_A-B_0)^2\nonumber\\
  &+\alpha_3^2(\mathbb{I}-cB_0-sY_AB_2)^2\nonumber\\
  &+\alpha_4^2(-cB_1+sX_AB_0+Z_AB_1)^2\nonumber\\
  &+\alpha_5^2(-cB_2+sY_AB_0+Z_AB_2)^2\nonumber\\
&+\alpha_6^2((\beta+\sqrt{4+\alpha^2})\mathbb{I}-I_{\alpha,\beta})^2\nonumber\\
  &+\alpha_7^2(X_AB_1-Y_AB_2)^2
\end{align}
where $c=\frac{\alpha}{\sqrt{4+\alpha^2}}$,$s=\frac{2}{\sqrt{4+\alpha^2}}$,$\alpha_6^2=\alpha_7^2=\frac{1}{4\beta}$,$\alpha_4^2=\alpha_5^2=\frac{\beta\sqrt{4+\alpha^2}}{2}\alpha_6^2=\frac{\sqrt{4+\alpha^2}}{8}$,$\alpha_1^2=\alpha_3^2=(\frac{\beta\sqrt{4+\alpha^2}}{2}-\frac{4+\alpha^2}{2})\alpha_6^2$, and $\alpha_2^2=\frac{\beta-\sqrt{4+\alpha^2}}{4}$. 
And the second one is,
\begin{align}\label{3-settingSOS2}
&(\beta+\sqrt{4+\alpha^2})\mathbb{I}- \hat{I}^{(2)}_{\alpha,\beta}\nonumber\\
&=\alpha_1^2(\mathbb{I}-cB_0-sX_AB_1)^2+\alpha_2^2(Z_A-B_0)^2\nonumber\\
  &+\alpha_3^2(\mathbb{I}-cB_0-sY_AB_2)^2\nonumber\\
&+\alpha_4^2((\Delta+s^2)B_0-(\Delta+1)Z_A+cZ_AB_0-csX_AB_1)^2\nonumber\\    &+\alpha_5^2((\Delta+s^2)B_0-(\Delta+1)Z_A+cZ_AB_0-csY_AB_2)^2\nonumber\\
     &+\alpha_6^2(-(\Delta+s^2)B_1+s(\Delta+1)X_A+\Delta cZ_AB_1-csX_AB_0))^2\nonumber\\
  &+\alpha_7^2(-(\Delta+s^2)B_2+s(\Delta+1)Y_A+\Delta cZ_AB_2-csY_AB_0))^2
\end{align}
where $c=\frac{\alpha}{\sqrt{4+\alpha^2}}$,$s=\frac{2}{\sqrt{4+\alpha^2}}$,$\alpha_6^2=\alpha_7^2=\frac{1}{4s\Delta(\Delta^2+s)}$,$\alpha_4^2=\alpha_5^2=\Delta\alpha_6^2=\frac{1}{4s(\Delta^2+s)}$,$\alpha_1^2=\alpha_3^2=\frac{1}{2S}-(\Delta+1)(\Delta+s^2)\alpha_6^2$, and $\alpha_2^2=\frac{\beta}{2}-\frac{\Delta^2+1}{s(\Delta+1)}$,and $\Delta=1$. }

{To make the SOS decomposition is positive semidefinite, it requires each $\alpha_i\geq 0$, thus  $\beta\geq\sqrt{4+\alpha^2}$.
And with some squared terms in (\ref{3-settingSOS1})-(\ref{3-settingSOS2}) acting on $\ket{\psi}$ are zero, it also can lead to the relations  (\ref{AP-relation1})-(\ref{AP-relation2}). Thus with the isometry given in the main text we can complete the proof of self-testing based on $ S_{\alpha,\beta}^{(2)}$. } 

%The maximum bound can be achieved by $B_0=Z,B_1=X,B_2=Y$ and $\ket{\Phi}=\cos \theta \ket{00}+\sin \theta \ket{11}$ with $\sin2\theta=\frac{2}{\sqrt{4+\alpha^2}}$. 

{For the first 3-setting steering inequality, three types of SOS decompostions can  be given,
the first one is,
\begin{align}\label{3z-settingSOS1}
&(\beta+\sqrt{4+\alpha^2})\mathbb{I}- \hat{I}^{(1)}_{\alpha,\beta}\nonumber\\
&=\frac{\beta}{2}(\mathbb{I}-Z_AB_0)^2\nonumber\\ &+\frac{\sqrt{\alpha^2+4}}{4}(\mathbb{I}-cZ_A-sX_AB_1)^2\nonumber\\&+\frac{\sqrt{\alpha^2-4}}{4}(\mathbb{I}-cZ_A-sY_AB_2)^2
\end{align}
the second one is,
\begin{align}\label{3z-settingSOS2}
&(\beta+\sqrt{4+\alpha^2})\mathbb{I}- \hat{I}^{(1)}_{\alpha,\beta}\nonumber\\
&=\alpha_1^2(-cX_A+sZ_AB_1+X_AB_0)^2\nonumber\\ 
     &+\alpha_2^2(-cY_A+sZ_AB_2+Y_AB_0)^2\nonumber\\     &+\alpha_3^2(S_Q\mathbb{I}-\hat{I}^{(1)}_{\alpha,\beta})^2
\end{align}
where $\alpha_1^2=\alpha_2^2=\frac{\alpha^2+\beta^2+\beta\sqrt{4+\alpha^3}+3}{4S_Q}$, $\alpha_3^2=\frac{1}{2S_Q}$.
and the third one is,
\begin{align}\label{3z-settingSOS3}
&(\beta+\sqrt{4+\alpha^2})\mathbb{I}- \hat{I}^{(1)}_{\alpha,\beta}\nonumber\\
&=\alpha_1^2((\Delta+s^2)Z_A-(\Delta+1)B_0+cZ_AB_0-csX_AB_1)^2\nonumber\\ 
     &+\alpha_2^2(-(\Delta+s^2)X_A+s(\Delta+1)B_1+\Delta cX_AB_0-csZ_AB_1))^2\nonumber\\
     &+\alpha_3^2((\Delta+s^2)Z_A-(\Delta+1)B_0+cY_AB_0-csZ_AB_2)^2\nonumber\\
  &+\alpha_4^2(-(\Delta+s^2)Y_A+s(\Delta+1)B_2+\Delta cY_AB_0-csZ_AB_2))^2
\end{align}
where $\alpha_1^2=\alpha_3^2=\frac{\beta}{4(\Delta+s^2)(\Delta+1)}$, $\alpha_2^2=\alpha_4^2=\frac{(1}{2s(\Delta+s^2)(\Delta+1)}$,$\Delta=\frac{\beta}{\sqrt{1+\alpha^2}}$.}

{
The PSD condition requires $\beta\geq0$. And with the first squared term in (\ref{3z-settingSOS1}) acting on $\ket{\psi}$ is zero ($\ket{\psi}$ is the state which maximally violates the steering inequality), it has the relations  (\ref{AP-relation1}), meanwhile with the linear combination of the second squared term in (\ref{3z-settingSOS2})and the first squared term in (\ref{3z-settingSOS3}) gives the relation (\ref{AP-relation2}). Thus with the isometry given in the main text we can complete the proof of self-testing based on $ S_{\alpha,\beta}^{(1)}$.
}

\textbf{\textit{self-testing for the measurements}}
{Above, we mainly focus on the states self-testing, for the  self-testing of the corresponding measurements (whose analysis can resort to \cite{Yang2014}) it will be similar. Starting with $ \Phi M_B(\ket{\psi})$ instead of  $ \Phi(\ket{\psi})$. Let's show it for one of the three measurements in 3-setting steering inequality cases for example.}

{After the isometry, the systems will be 
\begin{align}\label{eq:isometry2}
   \Phi(\underline{\tilde{Z}_B}\ket{\psi}) &=\frac{1}{4}[(I+Z_A)(I+\tilde{Z}_B)\underline{\tilde{Z}_B}\ket{\psi}\ket{00} \nonumber\\
    &+X_A(I+Z_A)(I-\tilde{Z}_B)\underline{\tilde{Z}_B}\ket{\psi}\ket{01} \nonumber \\
    &+\tilde{X}_B(I-Z_A)(I+\tilde{Z}_B)\underline{\tilde{Z}_B}\ket{\psi}\ket{10}\nonumber \\
    &+X_A\tilde{X}_B(I-Z_A)(I-\tilde{Z}_B)\underline{\tilde{Z}_B}\ket{\psi}\ket{11}]
\end{align}}

{With the relations  (\ref{AP-relation1})-(\ref{AP-relation2}) and the fact that $Z_AX_A=-X_AZ_A$, we find, $\tilde{Z}_B\tilde{X}_B\ket{\psi}=-\tilde{X}_B\tilde{Z}_B\ket{\psi}$.  By using this  anti-commutation relation between Bob's two measurements, one moves $\tilde{Z}_B$ to the left in the first, second, third and fourth lines  while
changing the sign of the forth line.
The analysis is then the same as the state self-testing, and the result is.
\begin{align}\Phi(\underline{\tilde{Z}_B}\ket{\psi})&=\ket{\text{junk}}[\cos\theta\ket{00}-\sin\theta\ket{11}] \nonumber\\
&=\ket{\text{junk}}[\underline{(I\otimes \sigma_z)}\cos\theta\ket{00}+\sin\theta\ket{11}]
\end{align}
Besides, from the SOS decomposition we can also find the relation  $\sin\theta Y_A(I+B_0)|\psi\rangle-\cos\theta B_2(I-Z_A)|\psi\rangle=0$. Thus  we have  $\tilde{Z}_B\tilde{Y}_B\ket{\psi}=-\tilde{Y}_B\tilde{Z}_B\ket{\psi}$. Following the above idea, we can finally conclude  the measurement in Bob's side are $B_0=Z,B_1=X,B_2=-Y$.}

\section{The transformation of a steering inequality into a  game}\label{appendixE}
In this section, we relate the constructed steering inequality to  a game which two party played to gain the score and build the relation between the quantum violation and success probability of the game defined. This is helpful for a direct comparison between different steering inequalities and it is necessary in the analysis of sample efficiency. For simplicity, here we only consider the 3-setting steering inequality.

In principle to obtain the maximum violation of the three setting steering inequality in main text Eq. (\ref{three-setting-I}),
the state between Alice and Bob should be $\cos\theta|00\rangle+\sin\theta|11\rangle$, which can be further written as 
$\frac{1}{\sqrt{2}}(\ket{\psi_0}\ket{+}+\ket{\psi_1}\ket{-})$, where we denote $|\psi_0\rangle=\cos({\theta})|0\rangle+\sin({\theta})|1\rangle$ and $|\psi_1\rangle=\cos({\theta})|0\rangle-\sin({\theta})|1\rangle$.

We define two   measurements in Alice's side $\{ |\psi_0\rangle,|\psi_0^{\dagger}\rangle; \;\; |\psi_1\rangle,|\psi_1^{\dagger}\rangle\}$, which actually the new measurements that introduced to substitute the measurements chosen in the main text in the real experiments. The measurements can also be written in the Pauli operators form,  $\{{A}_0={\cos({2\theta})\sigma_z+\sin({2\theta})\sigma_x};\;\;{A}_1={\cos({2\theta})\sigma_z-\sin({2\theta})\sigma_x}\}$.  

We notice that, if Bob gets $\ket{+}$, Alice takes${A}_0$,  Bob can conclude that Alice's qubit must be projected into $\ket{\psi_0}$; Meanwhile,  if Bob gets $\ket{-}$, Alice takes${A}_1$, then Bob  can conclude that Alice's qubit must be projected into $\ket{\psi_1}$. Since in steering scenario, Bob can sent information to Alice, such as the measurements result. Thus, this allows us to define the success probability of Bob guessing Alice's measurement result as,
\begin{equation}
    P_\text{virtual}^x=%p(A_0)\cdot
    p(A_0^0,B_1^0)+%p(A_1)\cdot 
    P(A_1^0,B_1^1),
\end{equation}
 which actually is related to the operators in the three setting steering inequality Eq. (34%\ref{three-setting-I}
 ).
More precisely, $\frac{\alpha}{2} Z+ XB_1=(\frac{\alpha}{2} Z+X)B_1^0+(\frac{\alpha}{2} Z-X)B_1^1=\frac{\sqrt{4+\alpha^2}}{2}(A_0B_1^0+A_1B_1^1)=\frac{\sqrt{4+\alpha^2}}{2}(2A_0^0B_1^0+2A_1^0B_1^1-I_B)$
 for $\sin(2\theta)=\frac{2}{\sqrt{4+\alpha^2}}$. Thus $P_\text{virtual}^x$ is related to  $\alpha\langle Z\rangle+\langle XB_1\rangle$. Similarly, we can define $P_\text{virtual}^y$ for $\sigma_y$ measurements scenario, which is related to  $\frac{\alpha}{2}\langle Z\rangle+\langle YB_2\rangle$.
 Together with the guessing probability for $(Z_A,B_0)$, we  define the total average passing probability as,
 \begin{equation}
    P_\text{virtual}=\frac{\sqrt{4+\alpha^2}(\frac{{P_\text{virtual}^x+P_\text{virtual}^y}}{2})+\beta p(a=b|Z_A,B_0)}{\sqrt{4+\alpha^2}+\beta}
\end{equation}
Thus we have,
 \begin{equation}
 \begin{split}
    P_\text{virtual}=\frac{\sqrt{4+\alpha^2}+\beta+S}{2(\sqrt{4+\alpha^2}+\beta)}=\frac{1}{2}+\frac{S}{2S_Q}
    %\frac{1}{2}+\frac{S-\sqrt{4+\alpha^2}}{2(S_Q+\sqrt{4+\alpha^2})}
   \end{split}.  
\end{equation}
This relation between the guessing probability and the violation holds for steering inequalities in Eqs.(30) %\eqref{eq:TCHSHsteering2_3} 
and (34) in main text.%\eqref{three-setting-I}. 
Thus the steering inequalities are transformed to testing games.

 \section{Robust self-testing of 3-setting inequality}\label{appendixF}
 In this section, we provide an analytical robustness bound for self-testing via 3-setting steering inequality. 
 
 We first consider the part of $G_1:=K_1- s (ZB_0+XB_1)-\tau_1 \mathbb{I}$ for $\mu \in (0,\pi/4]$, the spectral decomposition is already given in 
Eq. (\ref{CHSHdecom}).
To make $G_1\geq 0$, we consider the local extraction channel as, Bob takes $R_{1}=I$  with probability $q_1$, and takes $R_{2}=\sigma_z$  with probability $q_2$, meanwhile with the rest of the probability $1-q_1-q_2:=q_3$ Bob takes some other local extraction channel which  subjects to the choice of $B_2$. Then we have,
%\[
%\left\{
                \begin{align}
                 &q_1-s\lambda_1-\tau_1\geq 0 ;\nonumber \\
                  &q_2-s\lambda_2-\tau_1\geq 0 ;\nonumber \\
                   &s\lambda_{(1/2)}-\tau_1\geq 0 ;\nonumber \\
                 % s\lambda_2-\tau_1\geq 0 ;\\
                   &\Tr(\rho)=q_1+q_2+q_3=1;\nonumber \\
                  &\Tr(\rho \hat{B})=\lambda_1q_1+\lambda_2q_2+q_3\Tr(\rho YB_2)=S;\nonumber 
            \end{align}
           %   \right.
%\]
where $\tau_1=1-\gamma s$  with $\gamma\in[2,3]$. And $\tau_1$ should be less than zero.  We obtain $ s\geq\frac{1+q_3}{2\gamma-(\lambda_1+\lambda_2)} \geq\frac{1+q_3}{2\gamma-2\sqrt{2}}$.

Next, we  determine the value of $q_3$ to make $K_2$  is PSD. We notice $s\lambda_1-\tau_1$ and $s\lambda_2-\tau_1$ which according to the coefficients of $\ket{\psi_3}$ and$\ket{\psi_4}$ are greater than zero. That is, only the coefficients of $\ket{\psi_1}$ and$\ket{\psi_2}$ are greater than zero, $K_1$ part will be PSD. Therefore, we put$\ket{\psi_3}$ and$\ket{\psi_4}$ into $K_2$ part to make it PSD. Now $K_2$ part becomes,
\begin{equation}
\begin{array}{ll}
    G_2:=&q_3\Lambda^+_B(\psi_{1})+(s\lambda_1-\tau_1)\ketbra{\psi_3}{\psi_3}\\
    &+(s\lambda_2-\tau_1)\ketbra{\psi_4}{\psi_4}-sYB_2-(\gamma-3)s \mathbb{I}%\geq 0
    \end{array}
\end{equation}
 which is equivalent  to
 \begin{equation}
\begin{array}{ll}
    G_2 :=&q_3\Lambda^+_B(\psi_{1})+(s\lambda_1-\tau_1)\ketbra{\psi_3}{\psi_3}\\
    &+(s\lambda_2-\tau_1)\ketbra{\psi_4}{\psi_4}\\
    &-s(\gamma-2)(U\ketbra{\phi_1}{\phi_1}U^{T}+U\ketbra{\phi_2}{\phi_2}U^{T})\\
    &+s(4-\gamma)(U\ketbra{\phi_3}{\phi_3}U^{T}+U\ketbra{\phi_4}{\phi_4}U^{T})%\geq 0
    \end{array}
\end{equation}

where
 $ U=\left[\begin{smallmatrix}
  V  & 0  \\
 0  & V
\end{smallmatrix} \right]
\;\text{and}\;
       U^{T}=\left[\begin{smallmatrix}
     V^{*}  & 0  \\
 0  & V^{*}
    \end{smallmatrix}\right]  $
\begin{align}
  V=\left[\begin{matrix}
  \frac{ -\sin\mu_1i - \cos\mu_1\sin\mu_2)}{\sqrt{2-2\cos\mu_1\cos\mu_2}}& \frac{-\sin\mu_1i - \cos\mu_1\sin\mu_2}{\sqrt{2+2\cos\mu_1\cos\mu_2}}  \\
  \frac{\cos\mu_1\cos\mu_2 - 1}{\sqrt{2-2\cos\mu_1\cos\mu_2}}&   \frac{ \cos\mu_1\cos\mu_2 + 1 }{\sqrt{2+2\cos\mu_1\cos\mu_2}}  
\end{matrix} \right]
\end{align}
with $\phi_1=[\frac{ -1i}{\sqrt{2}},  0, \frac{1}{\sqrt{2}},   0],\phi_2=[   0, \frac{1i}{\sqrt{2}},  0, \frac{1}{\sqrt{2}}],
\phi_3=[  \frac{ 1i}{\sqrt{2}},  0, \frac{ 1}{\sqrt{2}},   0], $ and $
\phi_4=[   0, \frac{ -1i}{\sqrt{2}},  0,  \frac{ 1}{\sqrt{2}}]$.
The requirement of $ G_2\geq 0$ gives out,
                \begin{align}
        \frac{q_3(1+c)}{2}+(s\lambda_2-\tau_1)\text{overlap}^2({\psi_3},U^{-1}\phi_1)\nonumber\\+(s\lambda_1-\tau_1)\text{overlap}^2({\psi_4},U^{-1}\phi_1)-s(\gamma-2)\geq 0 ;\\                \frac{q_3(1-c)}{2}+(s\lambda_2-\tau_1)\text{overlap}^2({\psi_3},U^{-1}\phi_2)\nonumber\\+(s\lambda_1-\tau_1)\text{overlap}^2({\psi_4},U^{-1}\phi_2)-s(\gamma-2)\geq 0 ;        \end{align}
That is,
\[
\left\{                \begin{array}{ll}
                 \frac{q_3(1-c)}{2}+C_2\frac{\cos^2(\frac{\pi}{8})(\sin\mu_1-1)^2+\cos^2\mu_1\sin^2(\frac{\pi}{8}+\mu_2)}{4}\\+C_1\frac{\cos^2(\frac{\pi}{8})(\sin\mu_1+1)^2+\cos^2\mu_1\sin^2(\frac{\pi}{8}-\mu_2)}{4}-s(\gamma-2)\geq 0 ;\\
                 \frac{q_3(1+c)}{2}+C_2\frac{\sin^2(\frac{\pi}{8})(\sin\mu_1-1)^2+\cos^2\mu_1\cos^2(\frac{\pi}{8}+\mu_2)}{4}\\+C_1\frac{\sin^2(\frac{\pi}{8})(\sin\mu_1+1)^2+\cos^2\mu_1\cos^2(\frac{\pi}{8}-\mu_2)}{4}-s(\gamma-2)\geq 0 ;\\
                \end{array}
              \right.
\]
where $C_1=s\lambda_1-\tau_1$ and $C_2=s\lambda_2-\tau_1$.
With this channel, we have
$$\frac{q_3(1+c)}{2}+\frac{2-\sqrt{2}}{8}(s+\gamma s-1)-s(\gamma-2)\geq0$$ and
$$\frac{q_3(1-c)}{2}+\frac{2+\sqrt{2}}{8}(s+\gamma s-1)-s(\gamma-2)\geq0.$$

 It gives us $q_3c=\frac{\sqrt{2}}{4}(s+\gamma s-1)$
for $\gamma>2$ and $ q_3\geq \frac{-5\gamma + 2\sqrt{2} + 9}{-\gamma + 4 \sqrt{2} - 9}$. We can choose $\gamma=3$, which gives out $q_3=\frac{1}{2}$, and $s=\frac{3}{12-4\sqrt{2}}=0.4730$, $\tau=1-3s$. Thus we give the following robustness bound of one-sided self-testing based on three-setting steering inequality,
\begin{align}\label{CHSHsFidelity}
  F \geq s{S_{obs}}+\tau \geq \frac{3}{12-4\sqrt{2}}{(S_{obs}-3)}+1.
  \end{align}
 Although this does not reach the theoretical bound $s=\frac{1}{2(3-\sqrt{3})}$, the result is better than that of 2-setting inequality. This shows that adding more measurement settings can help to increase the robustness in one-sided self-testing.

\nocite{*}

\twocolumngrid

\end{document}